# HERITAGE: the concept of a giant flux neutron reflectometer for the exploration of 3-d structure of free-standing and solid interfaces in thin films.

**S. Mattauch[1], A. Ioffe[1], D. Lott[2], L. Bottyán[3], M. Markó[3], T. Veres[3], S. Sajti[3], J. Daillant[4], A. Menelle[5]**

[1] JCNS at MLZ, Forschungszentrum-Jülich GmbH, 85747 Garching, Germany

[2] Helmholtz Zentrum Geesthacht, 21502 Geesthacht, Germany

[3] Wigner Research Center for Physics 1525 Budapest, Hungary

[4] Synchrotron Soleil L'Orme des Merisiers Saint-Aubin - BP 48 91192 Gif-sur-Yvette, France

[5] Laboratoire Léon Brillouin CEA/CNRS, CEA Saclay, 91191 Gif sur Yvette, France

E-mail: s.mattauch@fz-juelich@de

**Abstract**. The instrumental concept of HERITAGE - a reflectometer with a horizontal sample geometry - fitted to the long pulse structure of a neutron source is presented. It is dedicated on creating a new class of reflectometers achieving the unprecedentedly high flux for classical specular reflectometry combined with off-specular reflectometry and grazing incidence small-angle scattering (GISANS), thus resulting in a complete 3-d exploration for lateral structures in thin films. This is achieved by specially designed neutron guides: in the horizontal direction (perpendicular to the scattering plane) it has an elliptic shape and focusses neutrons onto the sample. In the vertical direction it has a multichannel geometry providing a smooth divergence distribution at the sample while accepting the whole beam from a compact high-brilliance flat moderator.

The modular collimation setup of HERITAGE provides an extremely high flexibility in respect to sample geometries and environments, including the possibility to study all types of solid and liquid interfaces statically or kinetically. Moreover, the use of multiple beam illumination allows for reflectivity and GISANS measurements at liquid interfaces both from above and below without any movement of the sample.

This concept assures that reflectivity and GISANS measurements can be performed in its best way as the maximum possible and usable flux is delivered to the sample, outperforming all present-day or already planned for the ESS reflectometers and GISANS setups in flux and in measuring time for standard samples.

## 1. Introduction

The decision on the construction of the European Spallation Source (ESS) initiated a number of studies on the instrumentation for a long pulse neutron source. It will provide an exciting opportunity to design a reflectometer of the next generation to meet the increasing demand and anticipated scientific challenges [1]. In several meetings and in two specialized workshops carried out in 2012-2013, internationally recognized experts in the field of soft and hard matter discussed the science case for neutron reflectometer at the ESS. They identified the scientific drivers in which neutron reflectometry will assist in gaining valuable and unique information in the near future when the ESS will start operation. The anticipated research topics compromises a wide range of scientific disciplines, ranging from thin film magnetism and novel topological phases in confined geometries, over the functionality and properties of hybrid materials in the field of soft and hard matter to the structural biology of membrane proteins. A particular focus was put on the increasing complexity of thin film samples involving depth resolved two-dimensional patterning (exploration of 3-d structures in thin films) to enhance performance and creating new functionalities as well on the realization that nanoscale lateral structures of "natural" interfaces in soft matter are essential for their functionality.

Despite of the growing number of neutron reflectometers in the last decades at all neutron sources around the world and their steady advancements, neutron reflectometry research experiences up to today serious restrictions that handicaps the use of this technique for addressing the discussed science case in its full extent. The biggest drawback here is the generally low neutron flux available even at the nowadays strongest neutron sources that limits the accessible $Q$-range and thus the spatial resolution of the information gained from specular as well as off-specular neutron reflectivity. Another restriction consists in the low flexibility of the existing instruments that in general allows only the application of some particular modes at one experimental setup and instrument. If, for example, the 3-d structure of thin film samples should be investigated, it is desired to provide an easy switch between specular, off-specular and GISANS mode without being forced to move the sample. Latter is particular important for free standing liquids since any additional movement may influence the properties of the sample or require long waiting times for the relaxation to the initial state. Summarizing the requirements on a neutron reflectometer that overcomes the limitation imposed today, it should fulfil the following specifications: it should provide the highest possible intensity for GISANS and standard reflectivity mode and be built in the horizontal sample geometry to account for – beside other systems – the increasing demand for studies of free standing liquid interfaces. The concept should enable an easy and fast exchange between GISANS and standard reflectivity modes to increase the compatibility of the different measurement techniques carried out at one sample having been conditioned by certain external parameters. Flexible wavelength resolution from 1 to 10% is required for tuning the $Q$-resolution in respect to the scientific needs. For small length scales the $Q$-resolution will be relaxed to gain on flux, but for the investigations of thicker layers in the specular reflectivity case, for larger lateral structures on the scale of several hundred nanometres in the GISANS mode or to allow precise depth sensitivity in the TOF-GISANS mode, the $Q$-resolution needs to be adjustable. The wavelength band should be selected as wide as possible for the investigation of fast reactions and processes to probe e.g. kinetics of self-assembly of colloidal particles, folding of proteins or in-situ investigation of the exchange processes in membranes. Full polarisation capability, i.e. a polarised beam with subsequent polarisation analysis, is required for the measurements of magnetic samples. The polarisation capability may be also important for the soft matter samples to reduce incoherent background or to enhance the contrast by using a magnetic reference layer. The beam size needs to be optimized for typical sample sizes of 10x10mm$^2$ and 5x5 mm$^2$ that are typical for high quality samples produced by molecular beam epitaxy and pulsed laser deposition. Since in soft matter, but also in hard matter, samples will be available with larger surface areas, the instruments should also allow one to make use of the larger amount of material. Thanks to a new flat (also known as pancake) cold moderator [2] that will be installed at the ESS, the neutron beams will have 2.5-3 times higher brilliance in comparison with standard cold TDR moderators. As the result the peak intensity of the ESS will be about 60-75 times higher than the intensity of the time-of-flight instruments at the ILL. This opens an exciting opportunity to design a reflectometer of the next generation to meet the increasing demand and anticipated scientific challenges.

HERITAGE is an instrument concept dedicated to studies of 3-dimensional structures in thin films fulfilling the list of above requirements with best performance. Being designed for studies of both free-liquid and solid interfaces, HERITAGE possesses also all operation modes conceivable for a liquid reflectometer, including fast kinetic studies of liquid samples without the necessity of sample motion in the case of illumination from above and below.

The unprecedentedly high flux of HERITAGE achieved by using an optimized focussing elliptic neutron guide in the horizontal direction allows for studies of very thin, down to 5Å thick interfaces. This high flux can be traded for very high $Q_z$-resolution required for example for the depth-profiling that is currently only possible by X-rays and is a blind spot for neutrons due to limited intensity of present neutron instruments.

From other hand, the lateral length scales accessible by the multibeam focussing GISANS setup and off-specular reflectometry overlap well and permit the studies of lateral structures in an extremely wide range from 0.4nm to 30μm. All together this will allow for a gap-less exploration of the $Q$-space of thin interfaces in all 3 directions.

The full polarisation analysis in combination with the high flux will push the limits of signal to noise ratio for samples with inherent incoherent scattering, those studies were not possible before because of background issues.

With these characteristics HERITAGE takes a large share of the science case of ESS reflectometry. Apart from very small, below 1mm$^2$, samples the present HERITAGE design satisfies all foreseeable ambitions of the biological, hard matter and soft matter community.

## 2. General philosophy: relaxed $Q$-resolution machine

Investigations of thin interfaces (from several nm down to the sub nm range) require delivering as much usable intensity as possible to the sample position being able to access a dynamic range of 8 orders of magnitude and more (see Sect. 11). Fig. 1 shows the specular reflectivity curves simulated for a thin Fe-layer for perfect and relaxed $Q$-resolutions, respectively. Comparing the two cases it can be noticed that the resolution for the measurement can be drastically relaxed for thin interfacial structures without any loss of information. Indeed, with relaxed resolution the neutron intensity on the sample is significantly increased to obtain a detectable signal even from an extremely small amount of scattering material of a thin layer, particularly if it is required to measure the reflectivity up to high $Q$ values.

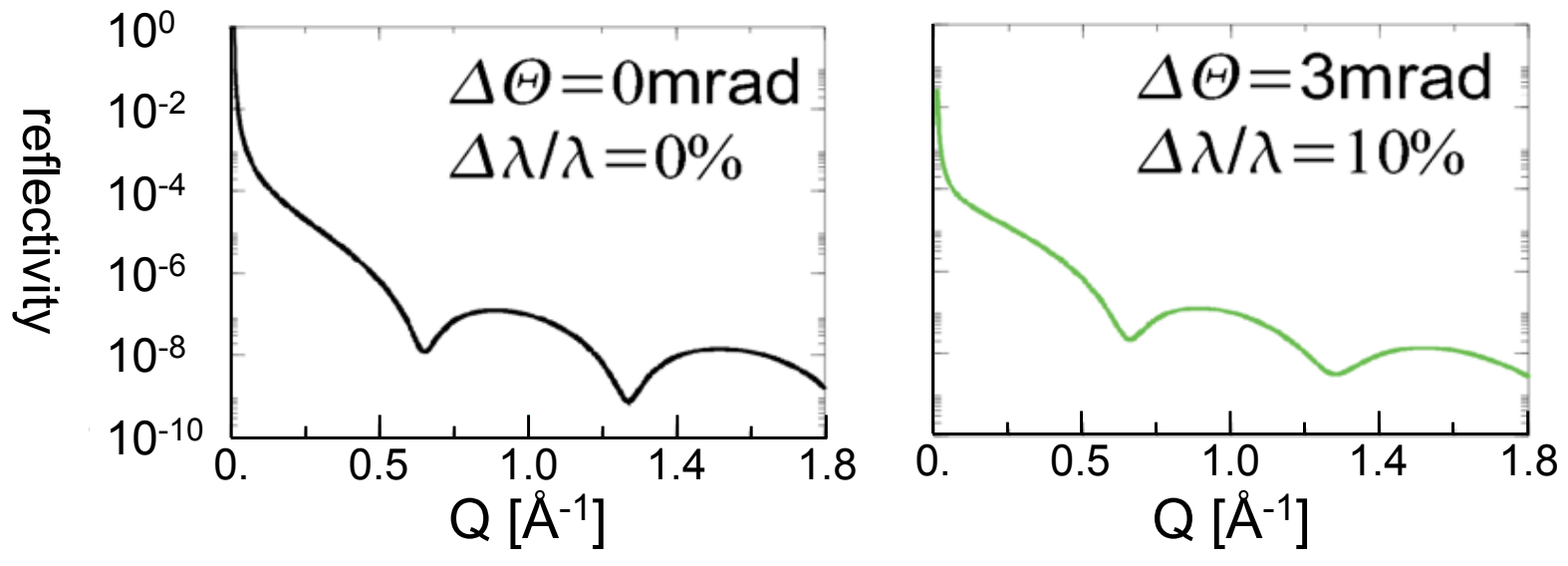

Fig. 1: Simulated reflectivity curves of 10Å thin Fe layer on an Ag substrate for ideal resolution/collimation (left) and for relaxed wavelength resolution/collimation (right) at 4.5Å.

The pulse width $\tau$ and the instrument length $L$ from the moderator to the detector are imposing physical limits on the main instrument parameters – the natural $Q$-resolution and the wavelength band $\Delta\lambda$, both are directly determined by the choice of $\tau$ and $L$:

$$\frac{\Delta Q}{Q} \propto \frac{\tau}{T} \propto \frac{\tau}{L} \qquad\qquad \Delta\lambda \propto \frac{1}{L} \tag{1}$$

($T$- flight time of the neutrons from the moderator to the detector). To achieve a resolution of 10% for $\tau$=2.86ms at the ESS and to use a maximum of neutrons from the spectrum centred around a wavelength of 3Å, the instrument length $L$ is fixed at about 36m. When $L$ is chosen no further relaxation of the wavelength resolution is possible, while the increase in the resolution can be achieved by artificially shortening the pulse length. In turn the repetition rate of 14Hz and the choice of instrument length defines the wavelength band of the instrument to $\Delta\lambda$=8Å.

## 3. General instrument layout

Recent moderator developments at the ESS resulted in the discovery of low-dimensional para-$H_2$ pancake moderators [3] that allow for a brilliance (i.e. the phase density) gain of up to an order of magnitude in comparison with a standard moderator. Such pancake moderators should certainly provide a significant intensity gain for neutron instruments with a high $Q$-resolution, as e.g. for small-angle scattering diffractometers and reflectometers that require a highly collimated beam (i.e. the selection of a small volume of the phase space).

For the ESS, where the low-dimensional moderator should provide neutrons for a number of neutron beams, its optimal form is a flat disk (pancake) with a height of about 3cm [2, 3]. Using a neutron guide with the similar height (3cm in the case of HERITAGE), however, results in an incomplete filling of the phase space that is accepted by the neutron guide and in a rather irregular divergence profile (see Fig. 2) at the exit of the 36m S-bended neutron guide. Such an irregular divergence profile is hardly usable for neutron reflectometry, where one relies on a smooth and flat divergence profile in the scattering plane.

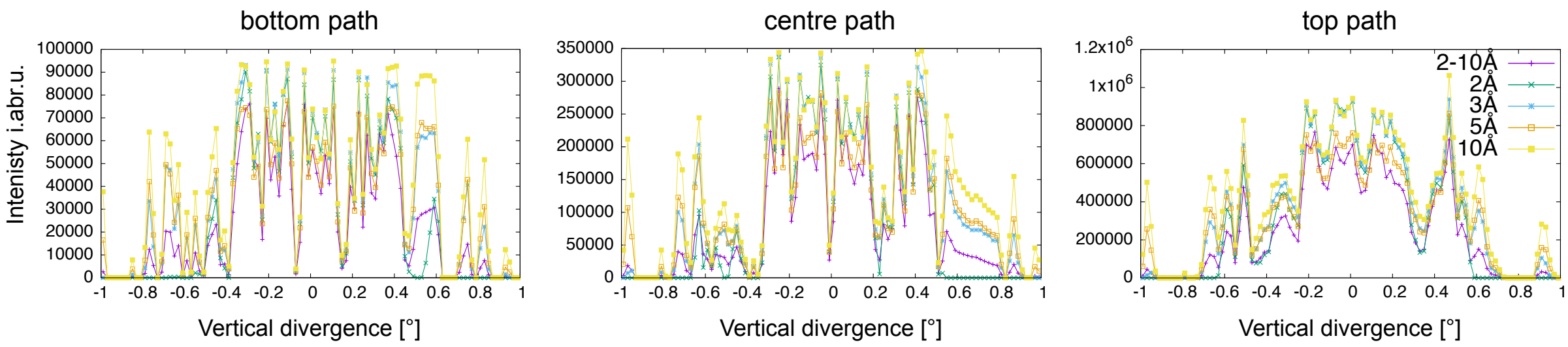

Fig. 2. Divergence profiles of 3 beams at the entrance of the "Liquid-Nose" (discussed in Sect. 4.1) for a non-channelled guide illuminated by a 3cm pancake moderator. The plots for the different wavelengths are scaled to each other.

This problem is here overcome by reducing the height of neutron guide by keeping it 3-5 times smaller than the height of the moderator. We achieve this by splitting the 3cm height neutron guide into 5 equidistant horizontal channels with heights of 6 mm each (Fig. 3, see also Table 1). Such a design results in a smooth divergence profile (see Fig. 9 in section 5) necessary for the use of the multibeam illumination system (Sect. 4). It furthermore allows one to take advantage of the enhanced brilliance of the flat moderator, while keeping a large total cross-section of the neutron guide in the flat direction and thus accepting a high neutron flux from the source.

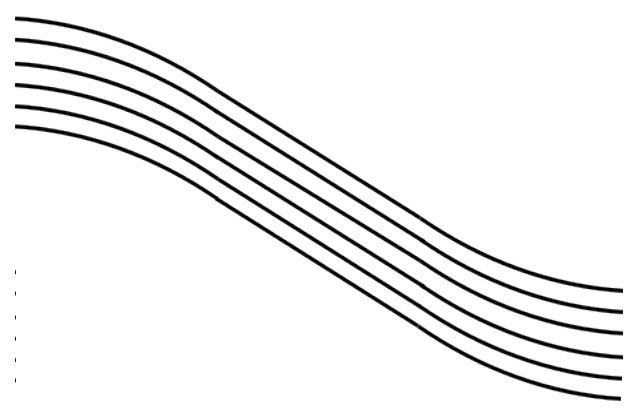

Fig. 3. Five channel S-guide in basic configuration.

The layout of HERITAGE is presented in Fig. 4. In the vertical plane the instrument guide is S-shaped in the vertical plane and made of two 10m long curved neutron guides ($R$=400m) with a 7m long straight neutron guide in the inflection point (see Fig. 3). The 2nd line of sight is lost at 24m distance from the moderator that is 10m before the sample position, so that the direct view of primary and secondary radiation sources is well avoided. The neutron guide is inclined by 2° against the floor to allow for studies of free liquid surfaces. The channel structure of the neutron guide will be interrupted in the centre of the straight part, 16m downstream of neutron beam, where the movable transmission polariser will be placed (details in Sect. 7).

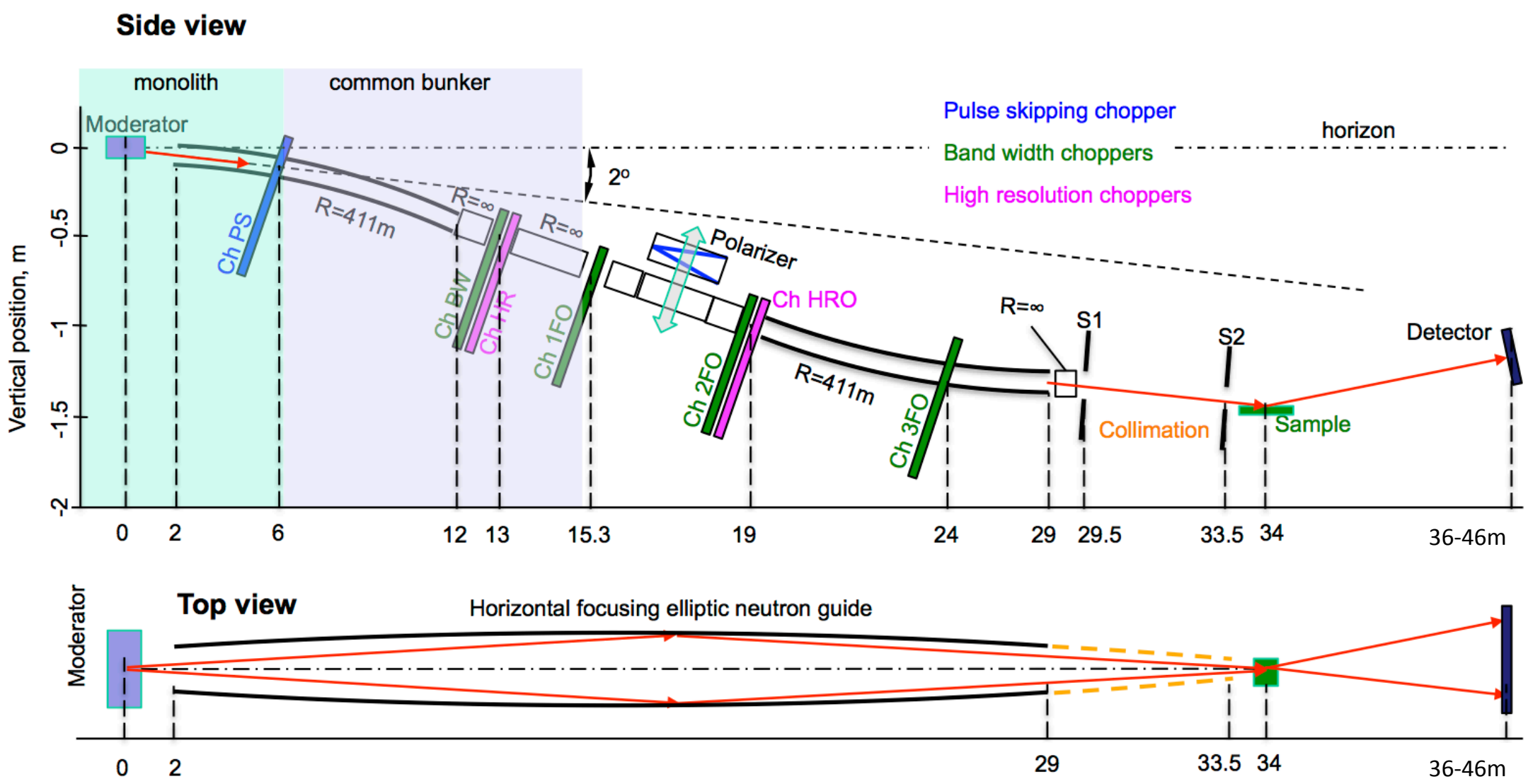

Fig. 4: The top and side view of HERITAGE. For clarity, the top view only shows the elliptic neutron guide. The end section of it (dashed brown lines) depicts a part of the collimation base and is exchangeable with other optical devices for different operational modes. The chopper design is given in [4].

In the vertical (scattering) plane the neutron guide ends 4.5m before the sample to provide a 4m long slot for the collimation options of the beam. In the horizontal plane, the elliptic neutron guide extends up to 0.5m in front of the sample. The last 4m of this section (shown by the brown dash lines in the top view of Fig. 4) can be modularly exchanged to provide different collimation options for HERITAGE operational modes described in Sect. 8. The basic guide parameters are listed in Table 1.

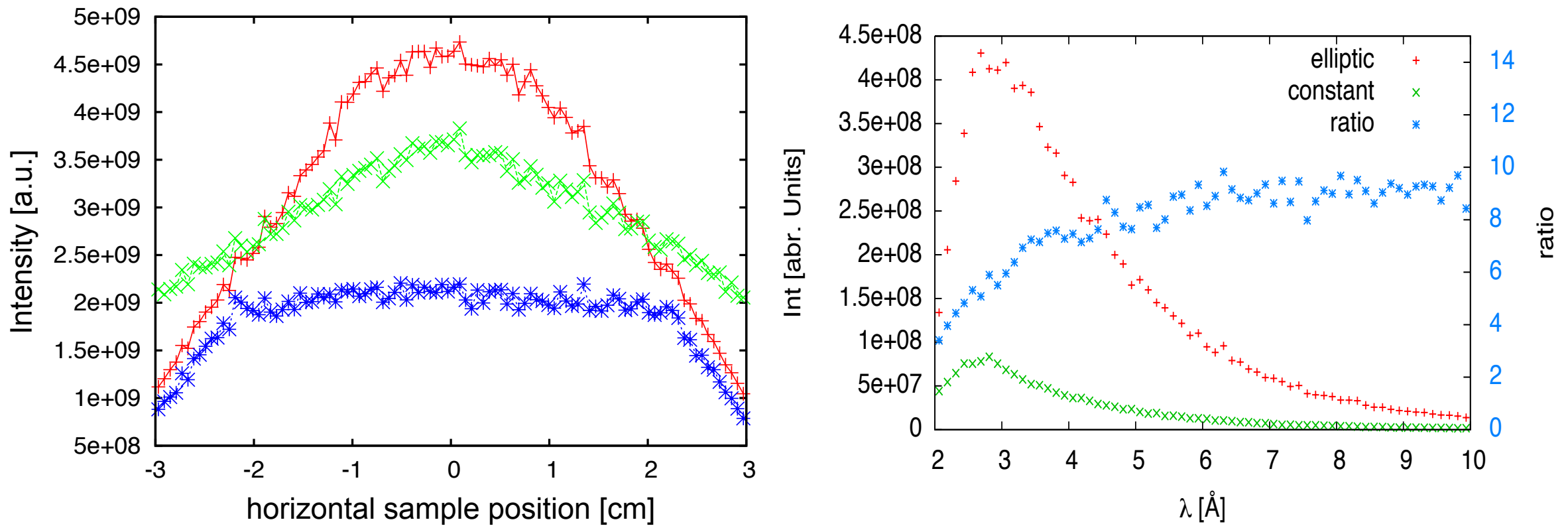

Fig. 5. Left: Effect of focussing in the intensity distribution on the sample. Red curve (+) - full focussing by 4m long end section of the elliptic guide, green curve (x) - focussing by 3m long end section, blue curve (*) - focussing in the absence of the 4m long end section, Red curve (+). Right: Comparison of the intensity at the sample position with a sample size of 1x1cm$^2$ for an elliptic and a straight neutron guide.

In the horizontal plane, the neutron guide has an elliptic shape with the focal points on the moderator and the sample. It allows one to focus a divergent neutron beam onto the sample in the horizontal plane (i.e. perpendicular to the vertical scattering plane). The ellipse is designed for the maximal transfer of brilliance from the cold source to samples with surface areas between 5x5mm$^2$ and 10x10mm$^2$. The focussing effect is demonstrated in Fig. 5, where the intensity distributions in the focal point at the sample position for different length of the end section of the elliptic guide (see Fig. 4, top view) are shown. For comparison the horizontal dashed line indicates the intensity distribution for the straight neutron guide would it substitute the elliptic guide. Another comparison between the intensities at the sample position (for a sample size of 10x10 mm$^2$) of this setup and the situation where the horizontal elliptic guide is replaced by a straight guide with the same coating, demonstrates a clear increase in the beam intensity by a factor of 4 in the case of elliptical focussing (see Fig. 5).

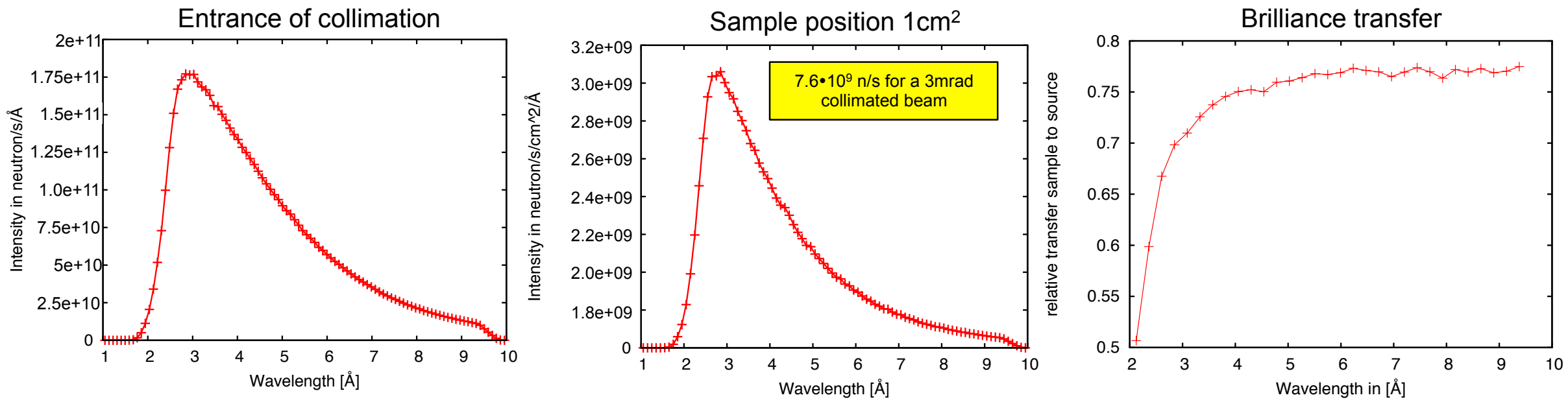

Fig. 6. Left: the intensity-wavelength distribution for the wavelength resolution of $\Delta\lambda/\lambda$=10% at the entrance of the collimation base (3x14cm$^2$). Middle: the spectrum of the beam collimated to 3mrad (Boxcar) in vertical direction on the sample. Right: wavelength dependence of the brilliance transfer of the whole neutron delivery system.

The performance of the neutron beam delivery system of HERITAGE is presented in Fig. 6. The left panel shows the intensity-wavelength distribution simulated for a low wavelength resolution of $\Delta\lambda/\lambda$=10% at the entrance of the collimation base. The spectrum of the beam formed by the collimation slits S1 and S2, both of 12mm height at 4m distance, i.e. at a beam collimation of 3mrad (Boxcar), is depicted in the middle panel. The integrated intensity over the full wavelength spectra amounts to $7.6 \cdot 10^9$ n/cm$^2$/sec ($5.4 \cdot 10^8$ n/cm$^2$ for each single pulse of the ESS).

The simulations above are supported by simple estimations: (i) compared with the ILL, the average flux at the ESS with the TDR moderator is about equal; the chopping of the beam will result in losses of the useful neutron intensity due to the blocking of the neutron beam between the neutron pulses by the choppers; (ii) the opening time is defined by the ratio of the pulse width to their period that is approximately equal to 1/25; (iii) such pulse structure is naturally produced by the ESS, so no losses related to the time structuring of the neutron beam will occur; (iv) the gain in flux because of the pancake moderator in comparison to the TDR moderator is about 2.5. Therefore, assuming that other beam parameters are similar, the expected gain is about 60. The actual gain may be even higher since a more advanced neutron optics will be used at a newly build reflectometer.

The brilliance transfer of the whole neutron delivery system is shown in the right panel of Fig. 6. It is calculated by the intensity ratio virtually measured by two identical brilliance monitors (using the VITESS simulation package) placed directly at the position of the moderator with a cross section of 3x10cm$^2$ (height x width) and at the sample position, respectively. Both monitors use an identical parameter set with the vertical and horizontal position set to 1x1cm$^2$ and the vertical divergence to ±0.9° (3mrad), as used in the intensity calculation in Fig. 6. The horizontal divergence is restricted to the maximum divergence accepted by the ellipse on the entry side to ±1.4°. The resulting brilliance transfer amounts to approximately 50% at $\lambda$=2Å and saturates at around 75% for $\lambda$>4Å.

As stated above, the last 4m of the horizontal elliptic neutron guide (sections 8-10 in Table 1) is exchangeable: the elliptic end used for standard reflectometry measurements at solid substrates can be straightforwardly substituted by other modules e.g. by a liquid nose allowing for measurements on free-liquid interfaces or a horizontal neutron collimation section to enable GISANS measurements. Such a modular design allows an extraordinary flexibility for the HERITAGE instrument with changes between the different instrumental operational modes in minutes. The modular setup is described in more detail in the following sections.

| Section | Length [m] | Radius [m] | channels | Entrance [mm] | | Exit [mm] | | Coating | | |
|---|---|---|---|---|---|---|---|---|---|---|
| | | | vertical | Width | Height | Width | Height | Left/Right | Bottom | Top |
| 1 | 10.0 | +400 | 5 | 95.8 | 30 | 194.5 | 30 | 3.0 | 3.0 | 40 |
| 2 | 1.0 | ∞ | 5 | 194.5 | 30 | 198.5 | 30 | 3.0 | 3.0 | 3.0 |
| 3 | 2.0 | ∞ | 5 | 198.5 | 30 | 202.7 | 30 | 3.0 | 3.0 | 3.0 |
| 4 | 4.0 | ∞ | 5 | 202.7 | 30 | 202.5 | 30 | 3.0 | 3.0 | 3.0 |
| 5 | 6.0 | -400 | 5 | 202.5 | 30 | 180.5 | 30 | 3.0 | 4.0 | 3.0 |
| 6 | 4.0 | -400 | 5 | 180.5 | 30 | 145.5 | 30 | 3.0 | 4.0 | 3.0 |
| 7 | 0.5 | ∞ | 5 | 145.5 | 30 | 140.0 | 30 | 3.0 | 3.0 | 3.0 |
| 8 | 2.0 | ∞ | 1 | 140.0 | 30 | 108.3 | 30 | 4.0 | - | |
| 9 | 1.0 | ∞ | 1 | 108.3 | 30 | 86.28 | 30 | 5.0 | - | |
| 10 | 1.0 | ∞ | 1 | 86.28 | 30 | 54.0 | 30 | 5.0 | - | |

Table 1. Basic guide design parameters: In order to avoid the depolarisation of the neutron beam, all guides after the polariser (section 4) are coated with non-magnetic super mirrors.

## 4. Chopper modes: general, higher wavelength resolutions (1%, 3% and 5%) and pulse skipping

The chopper positions are depicted in Fig. 4. The first chopper BW at 13m is the band selection chopper selecting an 8Å broad wavelength band. The band from 2 to 10Å will provide the highest intensity, however, the wavelength band can be selected arbitrarily in a range between 2 and 32Å. The additional frame overlap (FO) choppers 1FO at 15m, 2FO at 19m and 3FO at 25m prevent contaminations of neutrons with a wavelength of more than 50Å.

To allow for kinetic measurements (see Chapter 7) a pulse-skipping (PS) chopper is placed at 6.2m downstream of the moderator directly after the biological shielding of the target. In combination with the PS-chopper the general chopper settings enable one to have a time resolution of 140 ms, 210ms and 280ms by skipping 1, 2 or 3 pulses, respectively.

In order to achieve a high wavelength resolution for the proposed design, the natural pulse width has to be artificially reduced according to Eq. (1) in the wavelength frame multiplication (WFM) [8]. Technically the sub-pulse duration will be defined by the pulse-shaping chopper HR installed at 13m distance from the source. Each sub-pulse will provide neutrons arriving at the detector during a certain time interval (time sub-frame) covering a certain wavelength band (wavelength sub-frame) as shown in Fig. 7. For the optimal use of all neutrons from the long ESS pulse, a sequence of sub-pulses will be selected so that the wavelength band of subsequent sub-frames will overlap. Then the available wavelength band from 2 to 10Å will be completely covered without gaps in $Q$. To avoid ambiguity in the wavelength (i.e. in $Q$) determination the time sub-frames of these sub-pulses are well separated (see Fig. 7) by a high-resolution overlap (HRO) chopper at 19m.

Different resolutions require a variable opening of the pulse-shaping (high resolution, HR) chopper. Therefore a double disc chopper allows for different openings by setting an offset angle between the discs. It is important to note that in all three modes (1%, 3% and 5%), the $\Delta\lambda/\lambda$ value is not constant since the pulse length, whether directly from the ESS or from the HM-chopper, is always constant, while the $\lambda$ values change.

A detailed description of the chopper design including the pulse skipping mode is given in [4].

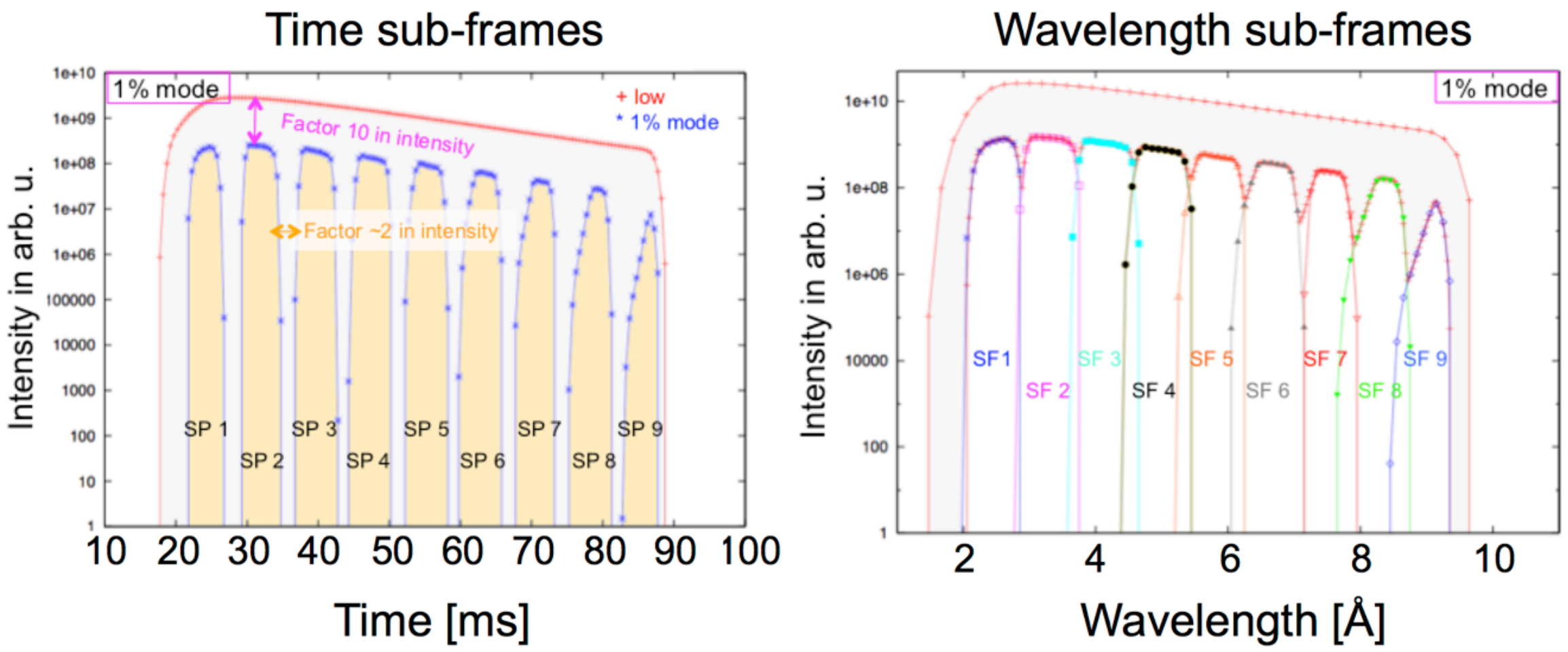

Fig. 7: The simulated time and wavelength intensity distributions for the 1% wavelength resolution modes. The low wavelength resolution curves ($\Delta\lambda/\lambda$=10%) are shown as envelops.

## 5. Multi-beam setup for fast kinetics and investigation of free liquid

### 5.1. Illumination from above

Wide $Q$-range reflectivity studies on free standing liquids in the TOF mode require 2-3 different incident angles onto the sample since the liquid surface itself cannot be titled against the horizontal. Latter is naturally possible for measurements on samples with solid-solid, gas-solid and liquid-solid interfaces.

It is furthermore important if not even essential to avoid any movement of the sample during and in between the experiment, as generated vibrations will result in the necessity of quite a long waiting time between successive measurements for the correct and accurate measurement of the reflectivity profile. In 1990$^{th}$ the idea of using multiple beams impinging on the sample under different incident angles for motion-free measurements at different $Q$ was realized at the V6 reflectometer [5] using the beams reflected from different crystals of vertically focussing monochromator and later at the REFSANS reflectometer [6] by bending the beam with the guide and reflecting mirror. Recently it was put forward as the basis for the design of the FREIA reflectometer proposed for the construction at the ESS [7].

For HERITAGE, another approach is suggested: here, three beams with the required inclination will be selected from a conventional neutron guide in the scattering plane by deflecting mirrors (labelled as “Liquid-Nose”) while an elliptic (horizontal) neutron guide will focus the neutrons onto the sample position perpendicular to the scattering plane. This allows HERITAGE to benefit from the enhancement in flux due to the focussing features of the elliptic neutron guide and deliver significantly more neutrons onto the sample position as by using the approach mentioned above. The horizontal focussing guide of HERITAGE will provide a significant gain factor of about 5 in respect to a straight guide solution as e.g. in the FREIA concept.

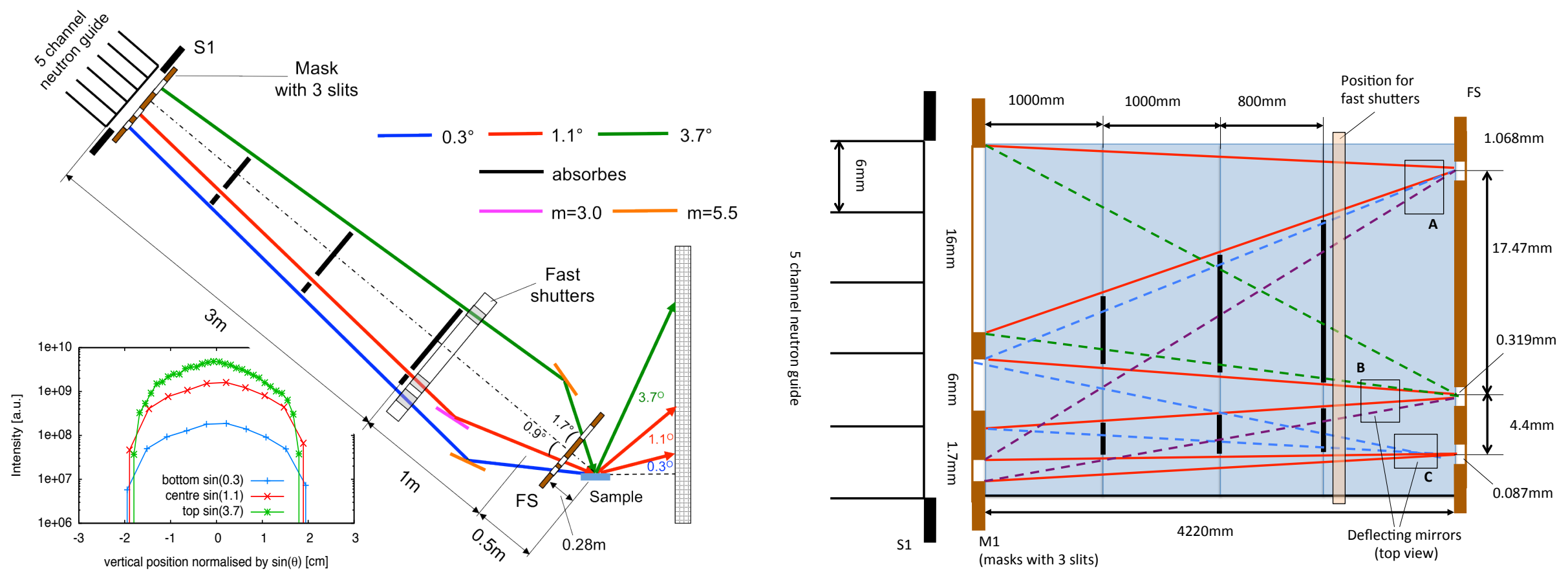

Fig. 8. Left panel: the sketch of the "Liquid-Nose". Blue, red and green lines depict the neutron beams impinging on the sample under incident angles of 0.3°, 1.1° and 3.7°, respectively. Black lines depict absorbers as shown in Fig.16, pink and orange lines the deflecting super mirrors with *m*=3 and *m*=5.5, respectively. Left panel inset: the positional intensity distribution on the sample. Right panel: The top view of the "Liquid-Nose". Red lines depict the "useful" neutron beams, blue, green and violet lines – parasitic beams that are blocked by absorbers shown by black lines.

The "Liquid-Nose" (see Fig. 8, left panel) setup is installed on a common support that can be moved in and out of the beam at the last part of the collimation base replacing a 1.2m long end section and part of the elliptic guide. Three beams illuminate the fixed horizontal sample under 0.3°, 1.1° and 3.7° and will be used in a multiplex mode, i.e. one after another. As the position and the deflecting angle of mirrors are slightly adjustable it is possible to select the most intense beam from the incoming divergence profile (see Fig. 9) in a very flexible way. The angular resolution and the footprint of the incident beam on the sample position are defined by the precision slits S1 and FS, respectively. The reflected intensity will be detected by a position-sensitive (PSD) detector or 3 independent single neutron counters. Three fast shutters will be installed in front of the deflecting mirrors that can be used for fast kinetic measurements on liquid interfaces. The cross-talk between single footprint slits in the FS is avoided by 3 neutron absorbing apertures as it is shown in Fig. 8 (right panel).

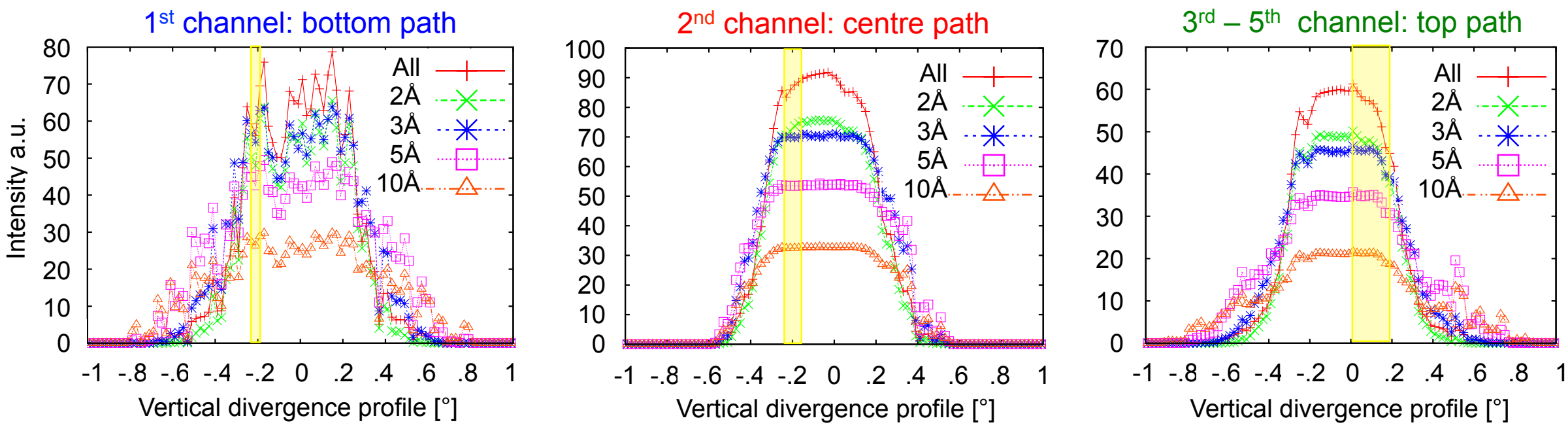

Fig. 9. Divergence profiles of 3 beams at the entrance of the "Liquid-Nose". Colour code corresponds to different wavelength in the spectrum. Yellow bars are indicating the angular width of vertically collimated beam selected by mask M1 and slits FS.

The *Q*-ranges corresponding to the angles 0.3°, 1.1° and 3.7° with a minimum and maximum wavelength of 2Å and 9.5Å, respectively, are well overlapping and enables one to access a *Q*-range from 0.007Å$^{-1}$ to 0.405 Å$^{-1}$ without moving the sample (Fig. 10 (left panel)). As an example, the neutrons delivered onto a 40x40mm$^2$ sample area at an incidence angle of 3.7° is calculated for the under-illumination conditions (right panel) and for an angular and wavelength resolutions of 4% and 10%, respectively. For the purpose of comparison, the slits and beam parameters are exactly the same as used in the ESS FREIA proposal [7] for the identical sample that allows for clearly demonstration of the gain factor of about 5 as expected due to the focussing effect of the horizontally focussing elliptic guide discussed above in Sect. 3 (see Fig. 5).

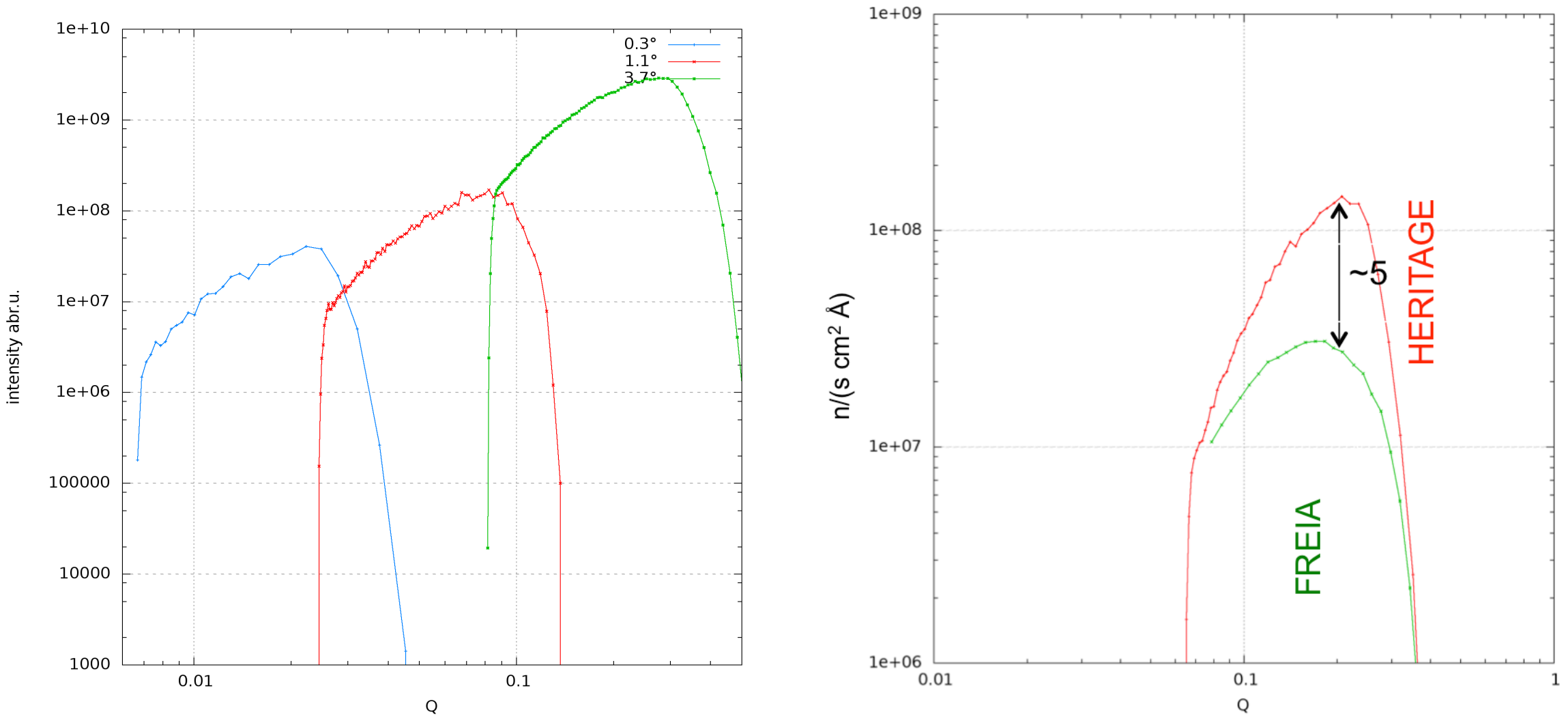

Fig. 10. Left: The $Q$-ranges covered by the incident angles 0.3°, 1.1° and 3.7°; the total accessible $Q$-range spans from 0.007$Å^{-1}$ to 0.405$Å^{-1}$. Right: the comparison of simulated intensity delivered by HERITAGE and FREIA at an under-illuminated sample of 40x40$mm^2$ for 4% angular and 10% wavelength resolutions for the incidence angle of 3.7°. The data for FREIA are taken from [7].

## 5.2. Illumination from below (inverted sample geometry)

In the ”illumination-from-below” operation mode free standing-liquid or liquid-liquid interfaces can be accessed from below the interface. The HERITAGE design provides a unique possibility to carry out such experiments without moving the sample during the measurements. For this purpose the “illumination-from-below” setup (Fig.11) is moved into the beam.

The beam deviated by 0.2° is selected and collimated by the slits S1 and S2. This beam is deflected upwards by one of 3 super mirrors (labelled as **A-A’, B** and **C)** creating beams incident on the sample at the angles $\theta_{in}$= 0.33°, 0.8° and 1.8°, respectively. The reflectivity curve will be measured subsequently in three steps, covering a total $Q$-range from 0.007$Å^{-1}$ – 0.197$Å^{-1}$ (Fig. 12) without vertical movement of the sample during the experiment. The $Q$-range can be even extended using the mirror A’’ installed 300 mm in front of the sample position that deflects the incident beam with an incident angle on the sample of $\theta_{in}$= 3.6°: In this way the $Q$-range is extended to 0.4$Å^{-1}$. As long as $m$=9 mirrors are not existing, it can be replaced by an m=7 mirror allowing for a maximum $Q$ value of 0.3$Å^{-1}$.

Using the “illumination-from-below” setup in the under-illumination-configuration (for details see section 10) given in Table 2 for a 40x40$mm^2$ large sample with an angular resolution of 4% will result into the intensity distribution on the samples as they are shown in Fig. 11 (right). It demonstrates that the intensity distribution along $Q$ for the individual angular settings possess a well-defined overlap. The slightly lower performance compared to the “Liquid-Nose” setup is accounted for the reduced focussing effect since the elliptic guide ends already in front of the collimation (see Fig. 5) and two mirrors are required for the **A**-configuration instead of one in the “Liquid-Nose” case. The right panel of Fig. 11 clearly shows that for all three configurations the neutrons hit the sample at the same position and no additional (vertical) movement of the sample is required.

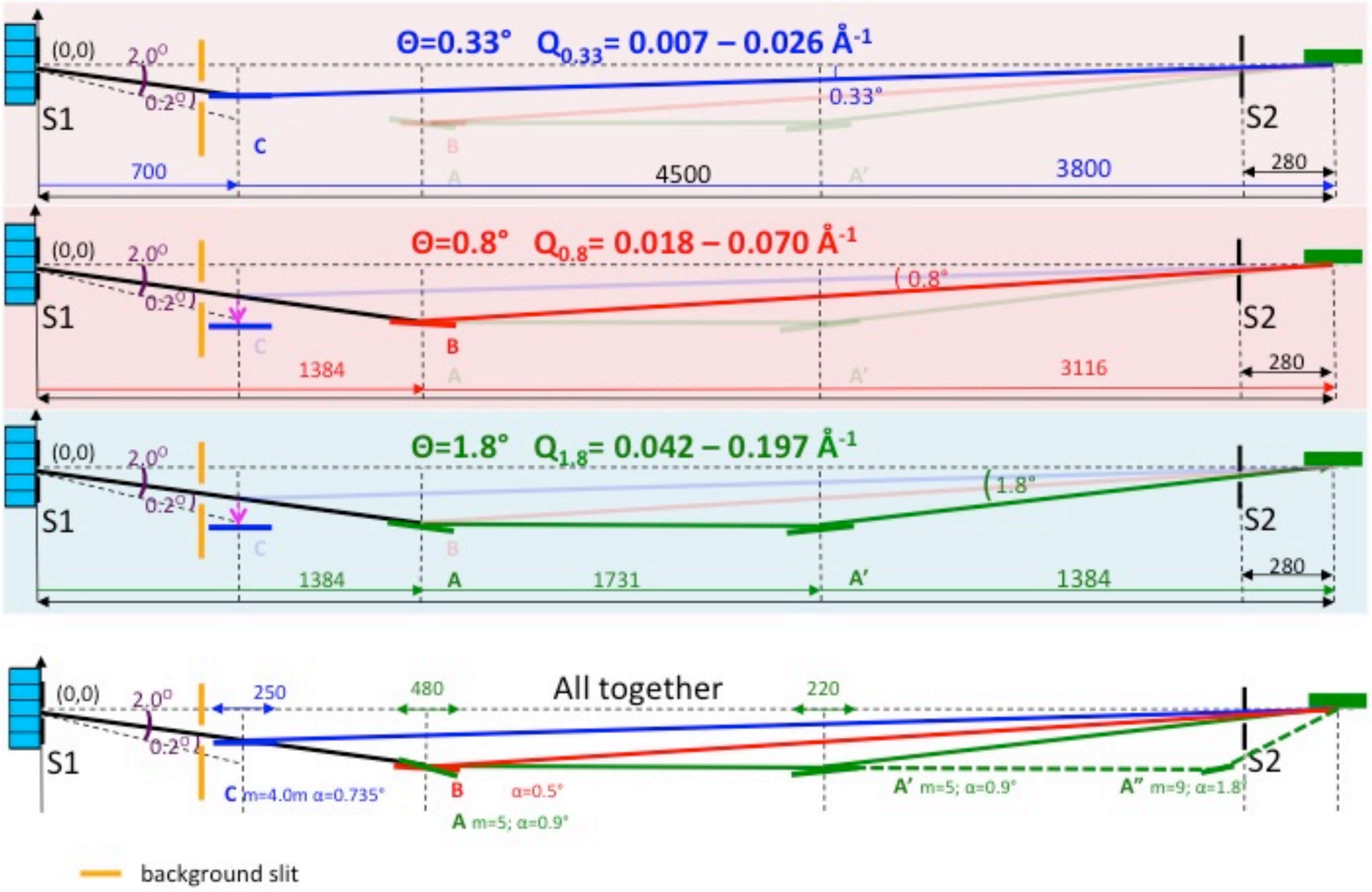

Fig. 11: Sketch for illuminating the sample from below. In the case of the 0.8° and 1.8° setup the mirror **C** will be moved out of the beam. Furthermore the mirrors **B** and **A** are identical so that just the inclination to the beam is changed. When using the configuration **B** or **A-A'** the mirror **C** is moved out.

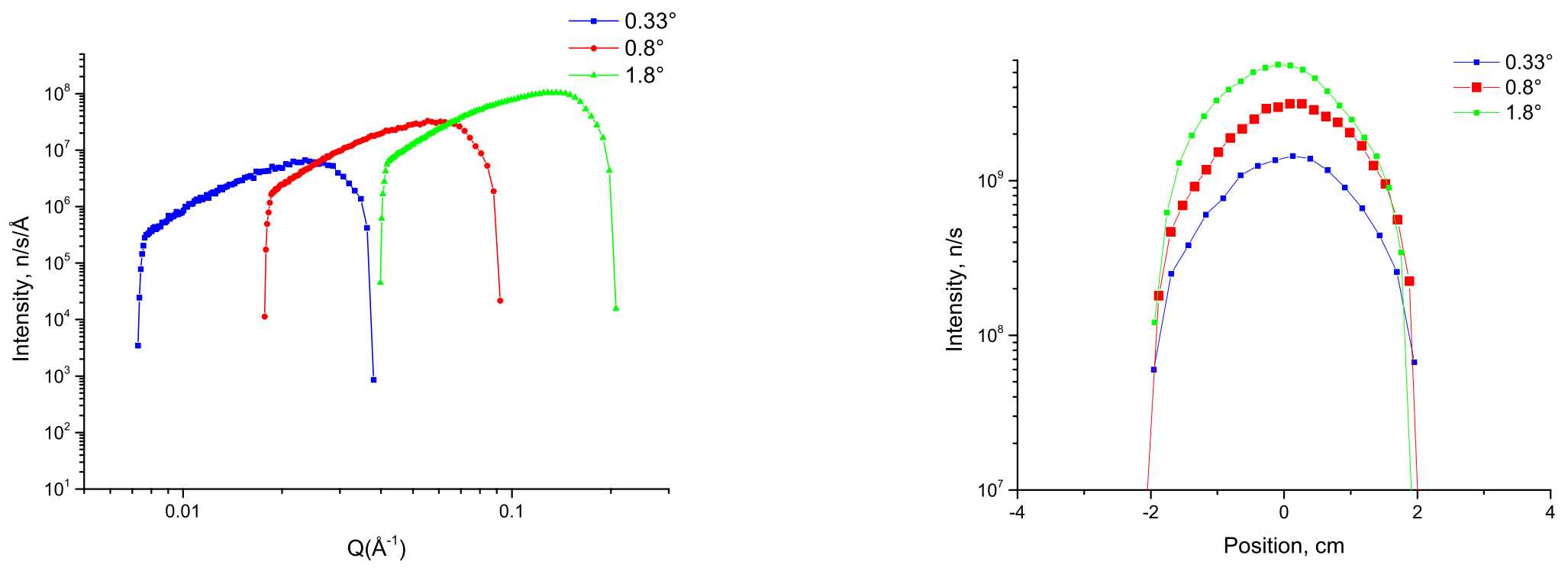

Fig. 12: Left panel: the *Q*-range overlap of the **A-**, **B-** and **C-**configuration. Right panel: the positional intensity distribution on the sample. Both simulations have been done in the under illumination configuration ($Q_{theta}$=4% resolution) for the slits settings shown in Table 2.

| θ, deg | S1 vertical | S1 horizontal | S2 vertical | S2 horizontal | S2-sample |
|---|---|---|---|---|---|
| **0.33** | 1.944 mm | 136 mm | 0.096 mm | 28.11 mm | 280 mm |
| **0.8** | 4.714 mm | 136 mm | 0.232 mm | 28.11 mm | 280 mm |
| **1.8** | 10.61 mm | 136 mm | 0.521 mm | 28.11 mm | 280 mm |

Table 2: Opening of the slits S1 and S2 in the vertical and horizontal direction for a 40 x 40 $mm^2$ sample ($Q_{theta}$=4% resolution) in the under illumination condition. The distance between S2 and the sample is 280mm and between S1 and S2 is 4220mm.

## 6. GISANS mode: the key element to the exploration of 3d-structures in thin films

The GISANS geometry requires a tight pinhole collimation not only in the vertical but also in the horizontal direction that results in a significant loss of incident intensity of about a factor 30 in respect of the reflectivity mode. The intensity, however, can be strongly increased by using a number of sub-beams focussing onto the same point in the detector plane (see Fig. 13) [6]. In this case the resolution of the GISANS setup will not be defined by the collimation of the incident beam as in conventional SANS, but as the ratio of the beam spot size $d_s$ in the detector plane to the sample-detector distance $L_{SD}$. The maximal $Q$-range of the GISANS setup is determined by the maximal detectable diffraction angle.

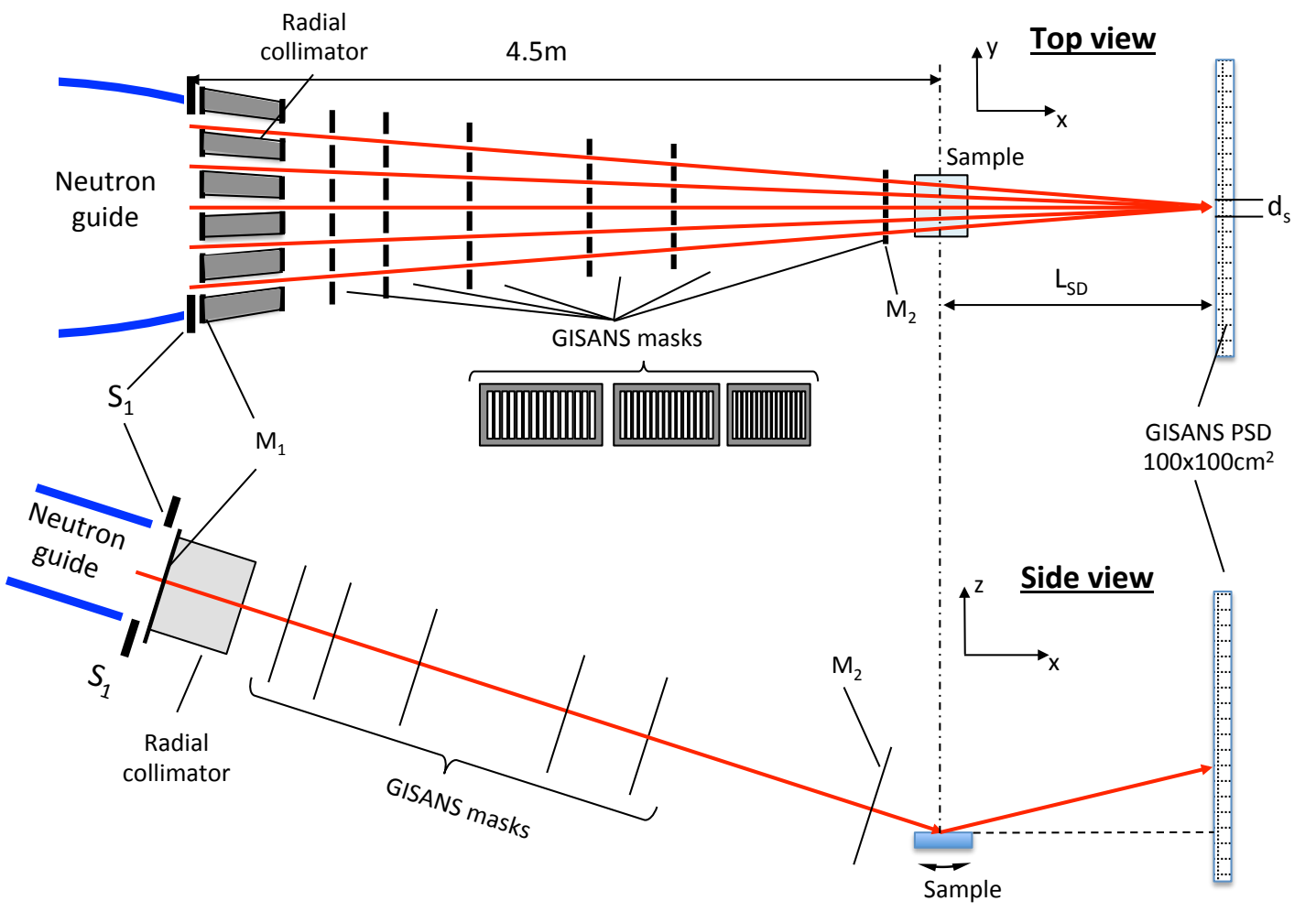

Fig. 13. Layout of the GISANS setup. Sub-beams propagating through comb-like masks $M_1$ and $M_2$ separated by 4.2m are focussed in the detector plane

The size of the focal spot $d_s$ (4mm, 8mm or 16mm) for different sample-detector distances given in Table 3 is defined by the vertical slit sizes of the masks $M_1$ and $M_2$ chosen for obtaining the required spot sizes $d_s$ in the detector plane (see Sect. 10). A crosstalk between sub-beams is prevented by a set of a radial collimator and six comb-like masks providing convergent neutron beam. They are placed inside of the collimation base replacing the complete 4m end section of the elliptic guide (Fig. 13). Mechanically, the positions of collimator and masks are fixed inside the modular unit and allow the GISANS option to be quickly moved in or taken out.

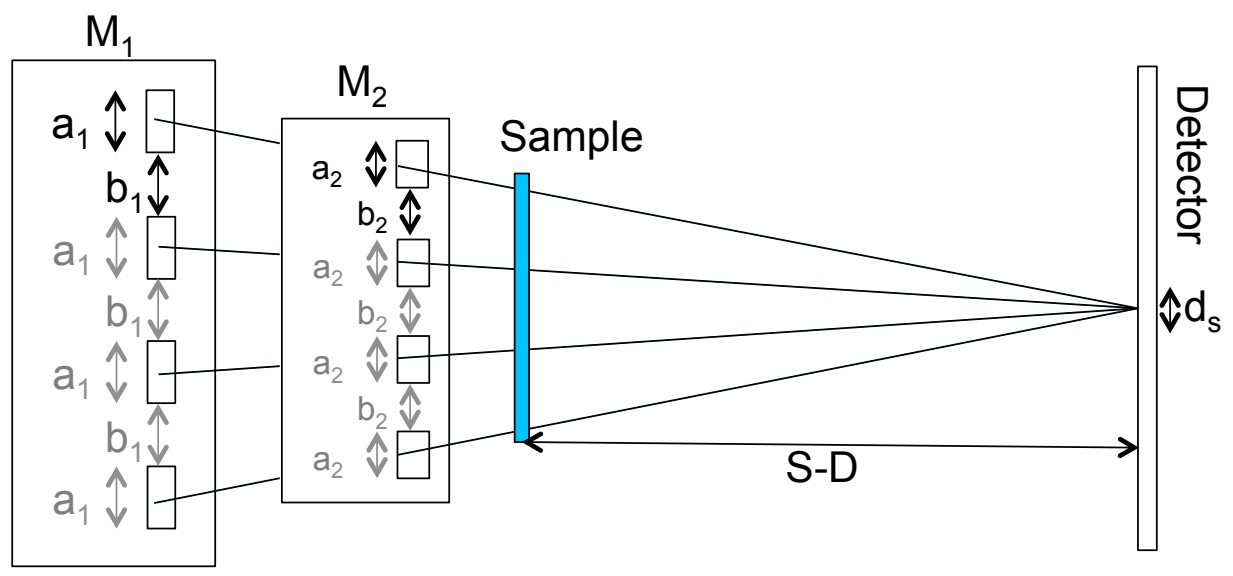

Fig. 14. Left: the geometry of the comb-like masks $M_1$ and $M_2$.

| S-D [m] | $d_s$ [mm] | $a_1$ [mm] | $b_1$ [mm] | $a_2$ [mm] | Number of sub-beams |
|---|---|---|---|---|---|
| 2 | 16 | 14,81 | 19,42 | 5,19 | 4 |
| 4 | 16 | 7,89 | 7,27 | 3,97 | 9 |
| 8 | 16 | 4,08 | 3,02 | 2,70 | 19 |
| 12 | 16 | 2,75 | 2,69 | 2,05 | 25 |
| 12 | 8 | 1,37 | 2,69 | 1,02 | 34 |
| 12 | 4 | 0,69 | 2,69 | 0,51 | 41 |

Table 3: parameters of slits for selected GISANS settings

A detector of a size $100\text{x}100\text{cm}^2$ and with a resolution better than $4\text{mm}^2$ will be installed inside of a vacuum detector tube and can be moved between 2m and 12m distance in respect to the sample position. One should note that the detector resolution (in the horizontal direction) should match the above calculated $d_s$ for the highest resolution or should be better by a multiple of integers.

The gain in the intensity due to the multi-beam arrangement can be estimated as the ratio between the integrated area of a single beam passing the mask $M_2$ and of the combined beam at the detector position resulting in a gain factor that is equal to the number of sub-beams focussed in the detector plane.

The performance of such a multi-beam setup in the case of the highest resolution ($d_s$=4mm, $L_{SD}$=12m) is illustrated by the simulations shown in Fig. 15. Plots at the upper panels demonstrate the effect of the focussing on the path to the detector. Each of the collimated sub-beams contributes equally to the spot at the detector position. The mask $M_1$ contains 41 slits of 0.69mm separated by 2.68mm, the mask $M_2$ contains 41 slits of 0.51mm separated by 2mm (Fig. 14). These slits provide an exact illumination of the $d_s$=4mm spot in the detector plane. The upper right plot clearly demonstrates that the chosen number of masks and their positioning avoid any significant cross-talk between the slits as no satellite peaks are observed.

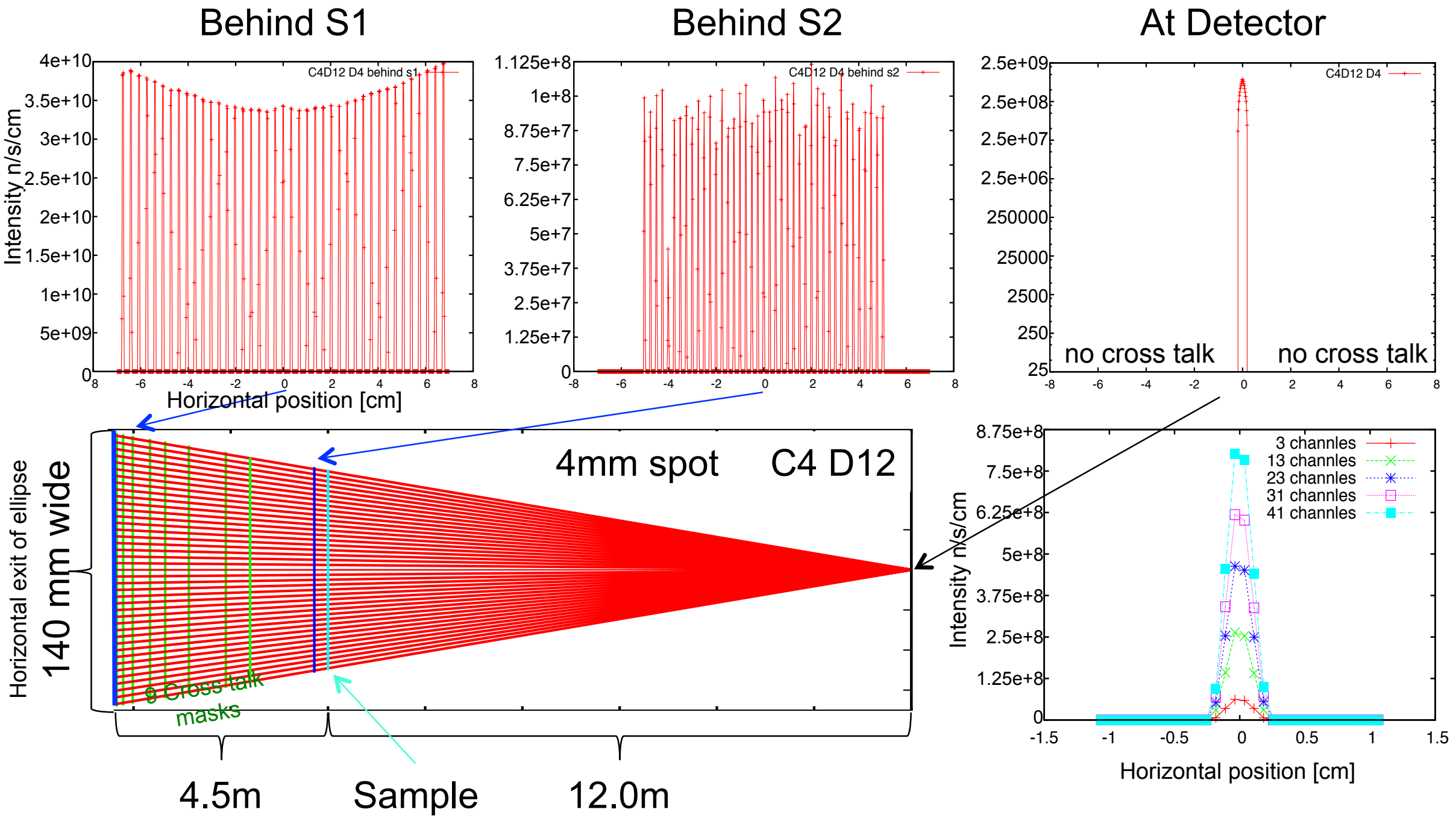

Fig. 15. Simulated performance of the GISANS setup for highest resolution of HERITAGE (4m long collimation length and 12m long sample-detector distance). The gaps in the middle upper plot are due to poor simulation statistics in the narrow channels.

The highest resolution of $Q_{y,}^{min}$=2.2·10$^{-4}$ Å$^{-1}$ is achieved by using 9.5Å wavelength neutrons for a 4mm spot size on the detector and at a sample-detector distance of 12m. The maximal $Q_y$ value of $Q_{y,}^{max}$ =1.5Å$^{-1}$ is accessed for 2Å wavelength neutrons and at a sample-detector distance of 2m. It enables one to probe an extremely wide range of length scales covering lateral structures from 0.5nm to 2.8μm (see Table 4). By accessing such large structures in the micrometer range the GISANS option at HERITAGE provides also a large overlap in the length scales accessible with off-specular scattering that can be very beneficial for the investigation of systems that inherit nanometer and micrometer objects (e.g. involving of magnetic domains on external parameters as magnetic field, etc.).

The large horizontal size of the neutron beam at the guide exit of about 140mm permits the application of a large number of sub-beams. As an example, for a beam spot of 4mm at the detector and a 12m sample-detector distances, about 40 sub-beams can be used for the GISANS measurements (see Fig. 15, right panel). Such a large number of sub-beams will also compensate for the intensity losses if e.g. a high angular resolution is required and result in an enormous intensity gain relative to a pinhole arrangement with the same resolution. Comparing to a hypothetical TOF GISANS instrument at the ILL providing an angular resolution of 4mm/12m=0.33mrad, the overall gain of intensity for HERITAGE can be rather conservatively estimated by multiplying the following factors: 25 due to the general intensity gain of the ESS in TOF mode, another 2 for the gain by the flat moderator vs. ILL cold moderator and a factor 40 due to the numbers of applicable sub-beams, thus resulting into an overall gain factor of 2000.

| Sample-detector distance | Size of the beam spot | $Q_y^{min}$ (9.5Å) - $Q_y^{max}$ (2Å) | Accessible lateral structures |
|---|---|---|---|
| 2m | 16mm | $5.3\times10^{-3}$ Å$^{-1}$ – 1.44Å$^{-1}$ | 0.4nm - 120nm |
| 4m | 16mm | $2.6\times10^{-3}$ Å$^{-1}$ – 0.77Å$^{-1}$ | 0.8nm - 240nm |
| 8m | 16mm | $1.3\times10^{-3}$ Å$^{-1}$ – 0.39Å$^{-1}$ | 1.6nm - 480nm |
| 12m | 16mm | $8.8\times10^{-4}$ Å$^{-1}$ – 0.26Å$^{-1}$ | 2.4nm - 710nm |
| 12m | 8mm | $4.4\times10^{-4}$ Å$^{-1}$ – 0.26Å$^{-1}$ | 2.4nm - 1420nm |
| 12m | 4mm | $2.2\times10^{-4}$ Å$^{-1}$ – 0.26Å$^{-1}$ | 2.4nm -2800nm |

Table 4: Accessible lateral structures in the GISANS mode for minimal and maximal sample-detector distance for detector size of 100cmx100cm. The minimal $Q_y$ is calculated from the $Q$-resolution, the maximal – for the detection at 100cm from the incident beam axis.

Increasing λ-resolution to 1% using the WFM mode [8] one can achieve high $Q$-resolution of about 1.5% by the cost of a factor of about 20 in respect to the low wavelength resolution mode [4] (see Sect. 3). Such a high $Q_z$-resolution allows for high-resolution depth profiling reflectivity measurements in thin films [9] (Fig. 17).

The remaining intensity gain of still 100 times over the existing SANS instruments opens up an extremely exciting opportunity for GISANS studies. Trading intensity against a drastically increased $Q_{x,y}$-resolution allows to push the upper limit in GISANS studies to 2.8μm (see Table 4) and to overlap the covered length scale by GISANS with the one by off-specular reflectivity measurements (1μm to 30μm). It closes the gap between the two methods (Fig. 16).

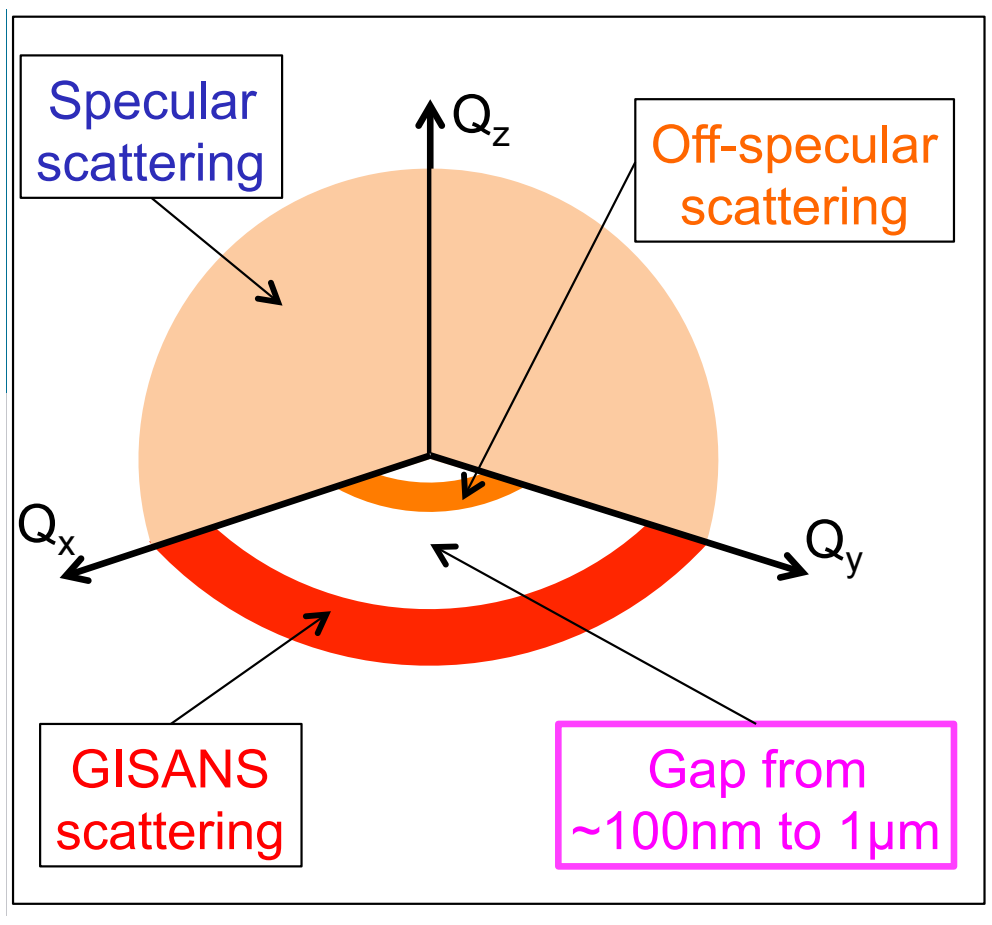

Fig. 16. Today's reflectometers are not able to close the gap between length scale covered by off-specular scattering and GISANS. With HERITAGE this gap will be closed allowing for a full 3d-exploration on thin films.

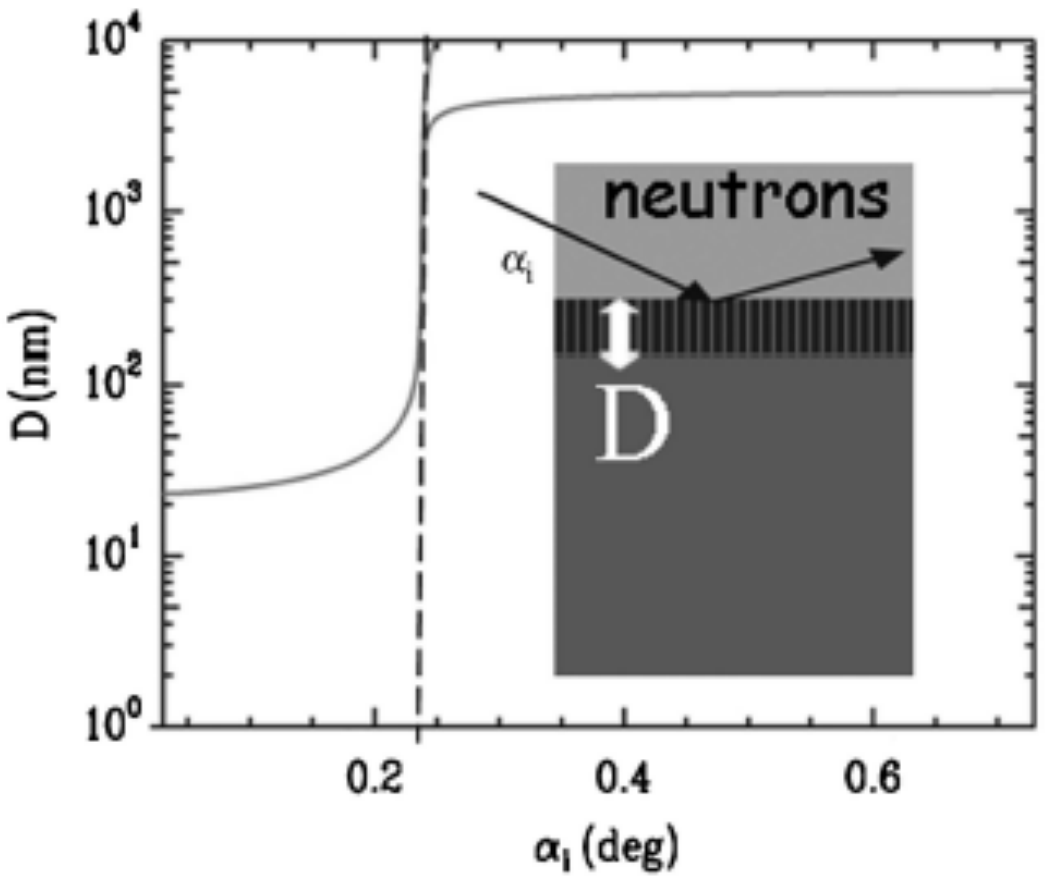

Fig. 17. Penetration depth $D$ as a function of the incident angle $\alpha_i$. The dashed line indicates critical angle $\alpha_c$. (Picture is taken from [9]).

Combining the GISANS investigations with a high depth resolution opens possibilities for studies of 3d-structures in thin films in a single experiment - an area that is up to today inaccessible for neutron research due to the limited intensity.

It should be noted that in order to take full advantage of the multi-beam focussing GISANS mode, the sample has to be sufficiently large to intercept all possible sub-beams. In the case of the highest $Q_y$ resolution ($L_{SD}$=12m) the optimum sample size is about 100mm, for a low $Q_y$ resolution ($L_{SD}$=2m) about 45mm. The reduction of the sample sizes will result in a proportional reduction of the number of the usable sub-beams and thus the incident intensity onto the sample for the GISANS measurement.

One should further note that the increase of the sample-detector distance from 2m to 12m results in the narrowing of the $Q_z$-range covered at a single angular setting by 20% (2Å to 7.5Å), so that the angular shift between angular settings has to be also reduced by 20%.

## 7. Kinetic mode

HERITAGE is ideally suited for the performance of kinetic measurements on samples of sizes 10x10mm$^2$ or larger due to the high intensity delivered at sample position. The kinetic mode is realized in two ways.

i) in the pulse skipping mode as it is described in the "Chopper modes" (Sect. 4)

ii) application of a fast shutters as it is proposed by the FREIA [7] proposal and the ESS

The first mode is basically best suited for solid based interfaces while the second option is particularly interesting for investigation of free standing liquid interfaces using the "Liquid-Nose" and its three beam geometry. For latter also the pulse-skipping mode is suitable and may be used as a backup solution if the fast shutters fail.

The pulse-skipping mode as well as the fast shutter solution can be easily combined with other modes as e.g. the high wavelength resolution mode or the GISANS mode and thus provides an extreme flexible experimental tool for the study of kinetics with neutrons in nearly any sample environment. The optimization on high flux allows for ultra-short measurement times. The simulation of such a kinetic experiment on HERITAGE is shown in Fig.18 for the air-$D_2O$ interface with a surface area of 40x40mm2 and the angular resolution of 4%. The slit settings for the "under-illumination-condition" are given in the Table 2.

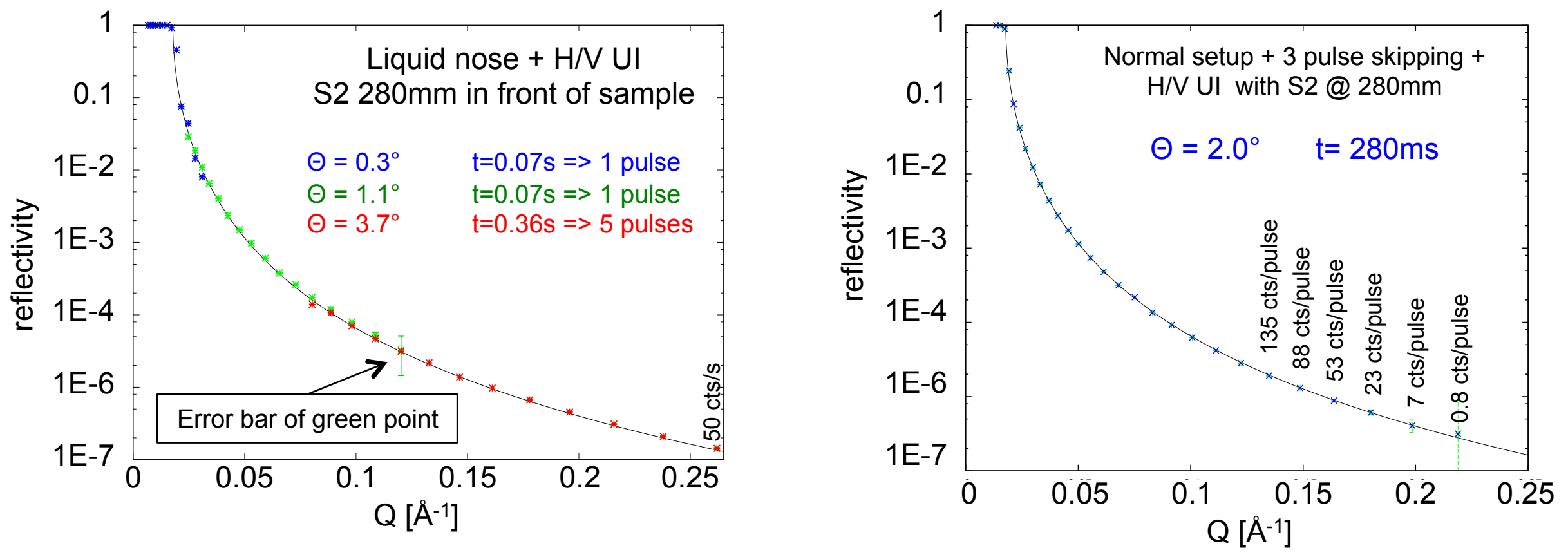

Fig. 18 : Simulated reflectivity from the 40 x 40 mm2 air-$D_2O$ interface. Left panel: reflectivity curve at HERITAGE measured by "Liquid-Nose" at 3 incident angles using a fast shutter system in 7 ESS pulses (in 0.5s). Right panel: the same curve measured by skipping three ESS pulses from a single ESS pulse at the incident angle of 2°.

In the fast shutter mode, a $Q_z$-range of up to 0.25Å$^{-1}$ is covered with all three angles in 11 pulses corresponding to a little more than one second total measurement time. At FIGARO [10] that is among the strongest reflectometer for the investigation of free-standing liquids at the present date, the measurement of a double sized $D_2O$ sample for the same $Q_z$-range required a total measurement time of 2 minutes, keeping in mind that some time is lost on Figaro for changing the guide configuration. Thus, the gain factor for HERITAGE is about two orders of magnitude. In the three pulse skipping mode a shorter $Q_z$-range up to roughly 0.18Å$^{-1}$ is covered with reasonable counting statistics, however, already in a measurement time of one pulse with a duration of 280ms. In order to extend the $Q_z$-range to the same range as it is available for the fast shutter mode, it would take about 20 sec for achieving reasonable statistics also up to high $Q$ values. It demonstrates clearly the advantage of the fast shutter approach over the one by skipping pulses for free standing liquids. On the other hand, in the case of a solid sample, the freely adjustable incident angle in the pulse skipping mode allows one to access nearly any $Q$ value at which the kinetic process can be accessed. In such a way both methods are complementary and enable one to study fast kinetics around low and high $Q$ values.

## 8. Polarisation and polarisation analysis

In the polarised mode, the central 1m long piece of the 5-channel guide will be replaced by a double channel polarising cavity (Fig. 19), which is built upon thin, 0.3mm thick, Si wafers coated with $m$=5 Fe/Si super mirrors working in transmission. This solution leads to very high values of neutron beam polarisation (see Fig. 20) with small intensity losses over the whole wavelength band and allows for the practically instant switch between polarising and non-polarising operation modes without affecting the overall beam propagation. Even higher polarisation of the neutron beam of more than 99% can be achieved using an optional $^3$He neutron spin filter, however, by the price of a reduced beam intensity by 25%. It should be noted that all sections after the polariser are coated by non-magnetic super mirrors.

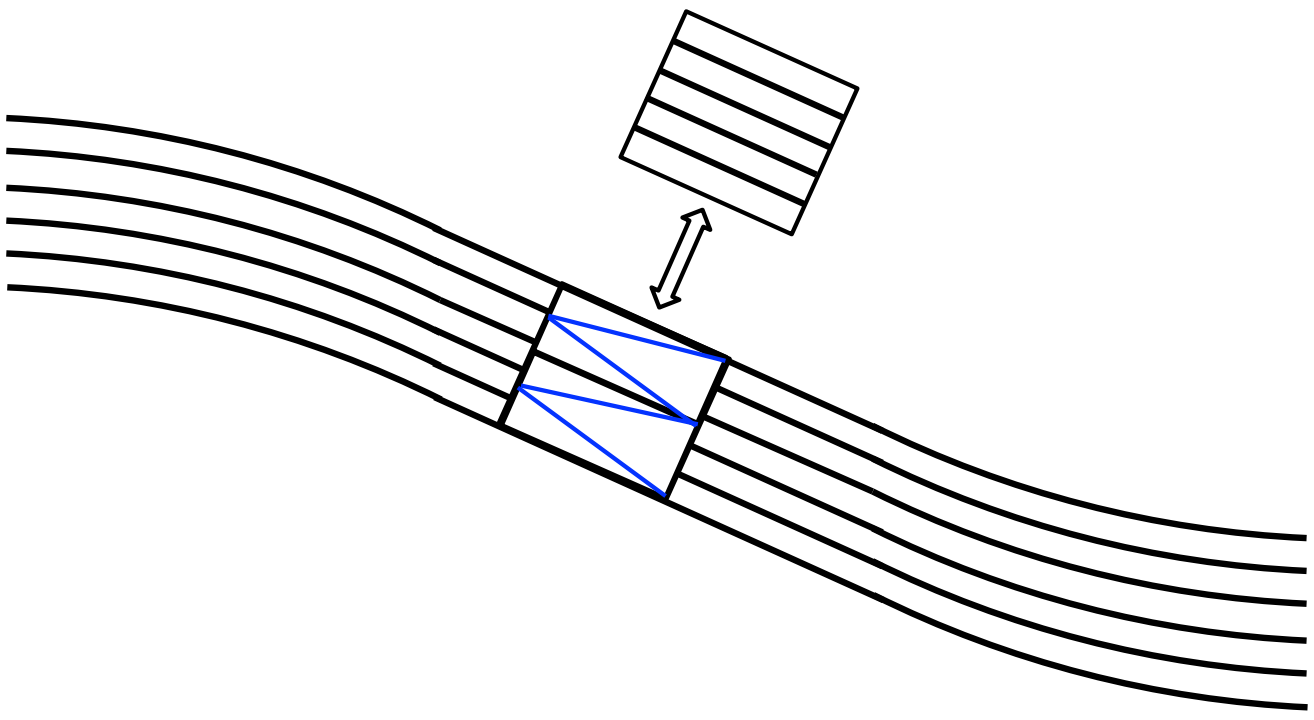

Fig. 19: Top view of the S-guide with the polarising cavity exchangeable with a non-polarising neutron guide section.

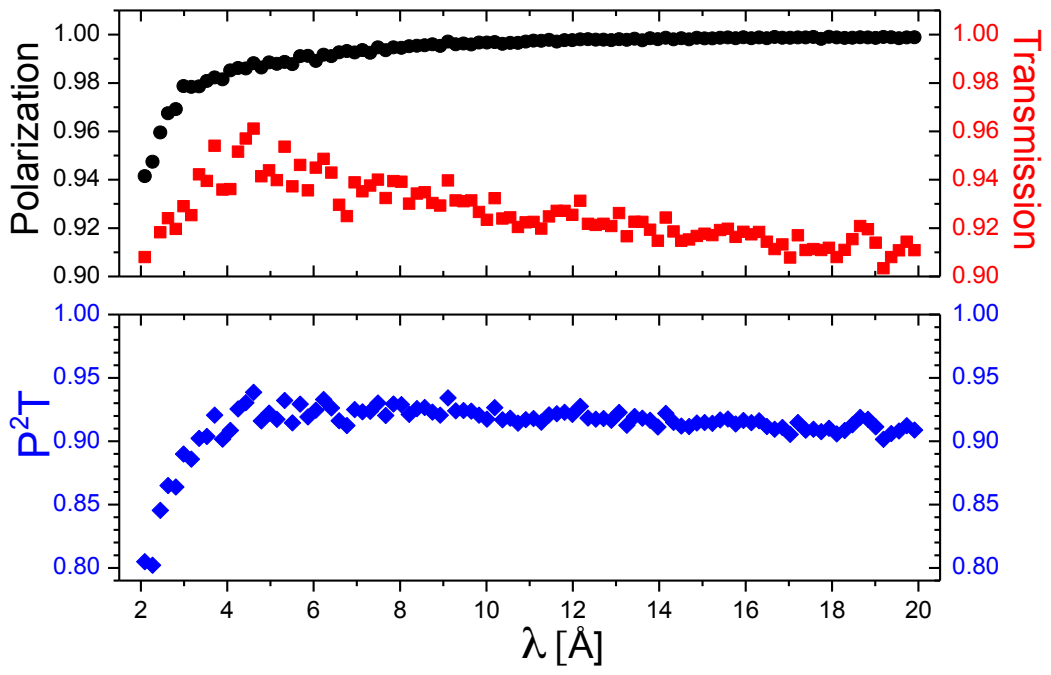

Fig. 20. Polarisation P, transmission T of the selected spin component and the figure of merit $P^2T$ of the polarising cavity setup.

For the polarisation analysis a polarisation analyser can be placed between the sample table and the detector and can be quickly moved in or out. Different analyser types are foreseen for the different operation modes:

- For the specular reflectivity mode, when the detected beam has a rather small horizontal cross-section and angular divergence, a transmission analyser made of a double-side super mirror coated Si wafers will be used. In front of it an adiabatic RF-flipper providing extremely high flipping efficiency over the whole wavelength band will be installed.
- For the off-specular reflectivity and GISANS modes, when polarisation analysis should be performed over a wide solid angle covering a large area of a position-sensitive detector, $^3$He neutron spin filters will be used. The use of the on-beam optically pumped large SEOP cells is here envisaged, where a very high degree of the polarisation (>80%) is already achieved today, resulting into a high transmission and polarisation efficiency that can be provided over the whole wavelength band of about 8Å [11, 12]. Such a solution will ensure that the divergence of the scattered beam will be kept unchanged during the polarisation analysis. Moreover, the well-developed adiabatic fast passage technique allows for very effective flipping of $^3$He spin states, thus no additional flipper after the sample is required.

It should be noted that a polarisation analyser will be not only essential for magnetic studies with the reflectometer, but will also help to reduce significantly the incoherent background scattering from hydrogen in soft matter samples.

## 9. Modularity

All the modes described above can be freely and easily combined as depicted in Figure 20. Additionally, the flexible wavelength resolution mode (10%, 5%, 3% and 1%), the polarisation and polarisation analysis, as well as the two kinetic modes (pulse skipping and fast shutters) can be combined with the reflectometry mode (specular and off-specular) or the GISANS mode and even with the special liquid sample setup ("Liquid-Nose" and "illumination-from-below" modes) without any restriction. More than 20 different modes are thus feasible, each easily accessible and quickly alterable in a few seconds.

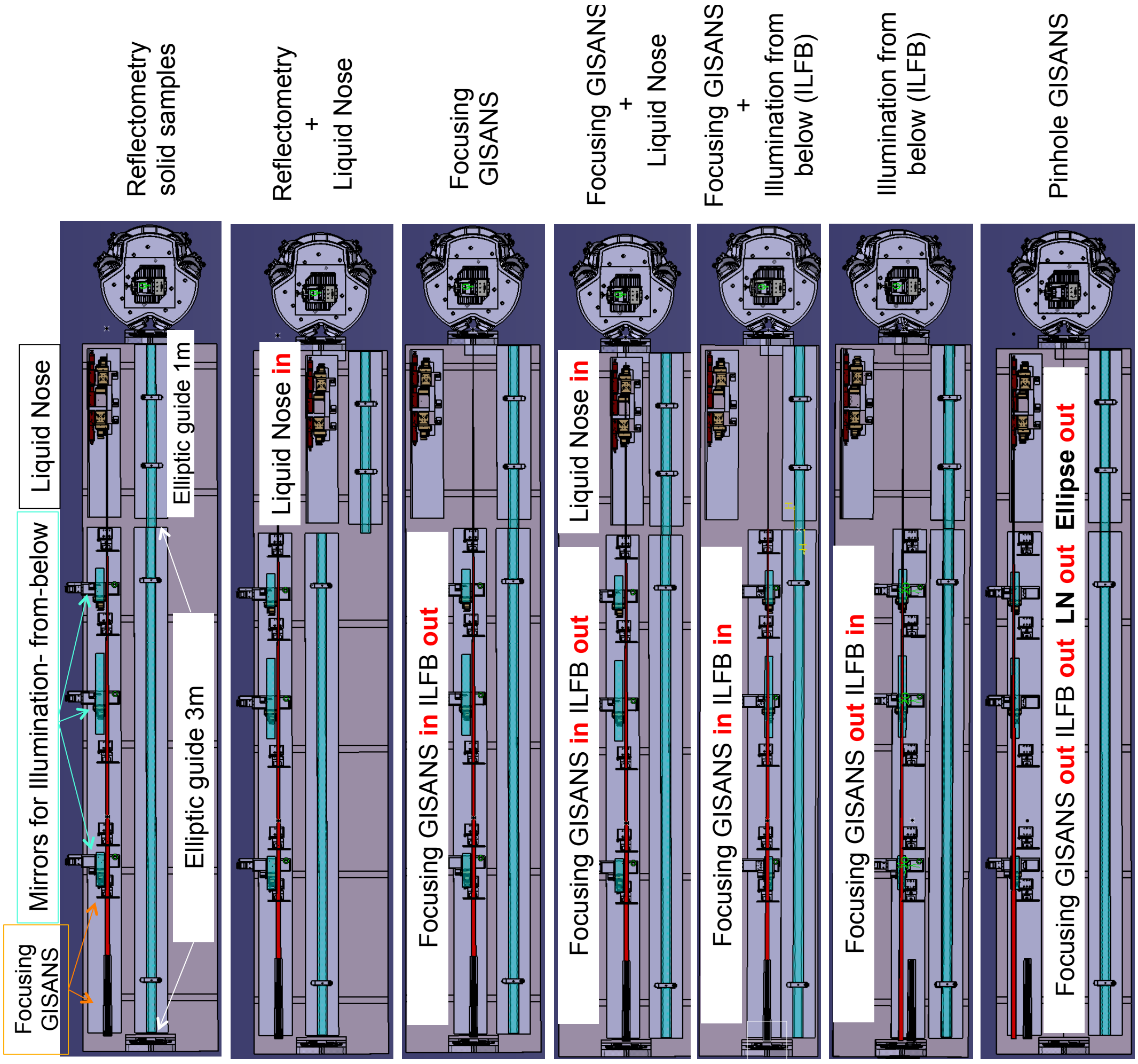

Fig. 20. A sketch illustrating the modularity of the different operational modes of HERITAGE. Optical elements – "Liquid-Nose", "GISANS", "Ellipse" and "Illumination from below" – move in and out sharing the place in the collimation base.

## 10. Under-illumination of the sample

Avoiding any illumination of areas beyond the pure sample area in the sample plane, called the under-illumination of the sample, is extremely important for reflectometry and GISANS experiments since scattering from sample holders is one of major sources of background that is very hard to separate from the actual signal. The footprint slit FS controls the illumination of the sample by placing it at the distance $L_s$ from the sample (Fig. 21).

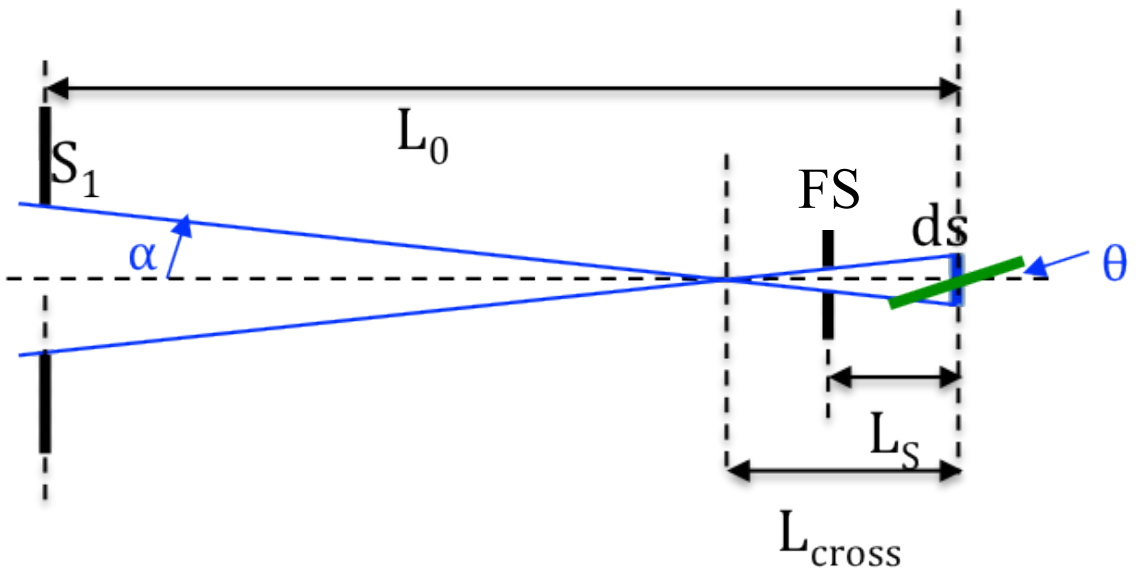

Fig 21: Slit geometry (side view) for under illumination of the sample (shown in green) with the effective size *ds*. For horizontal direction the geometry is the same with *ds* equal to the real width of the sample.

Obviously the position $L_s$ of the slit FS defines its size and therefore the intensity at the sample. The closer the slit FS is to the sample the less intensity is lost by the under-illumination condition. The minimal distance $L_s$ to the sample is defined by the size of the sample environment around the sample. For hard matter samples that are also studied at low temperatures and strong magnetic fields, closed cycle cryostats are commonly used nowadays. A typical cryostat in which a 10mm x 10mm sample can be easily placed has an outer radius of about 25mm. For liquid samples, on the other hand, rather bulky environment are often necessary. For purposes of simulations demonstrated in this article, $L_s$=25mm for 10mm x 10mm samples and $L_s$=280mm for liquid samples with size 40mm x 40mm are considered (Table 7).

## 11. Measurements of extremely small reflectivities (up to $10^{-9}$)

The gain factor of more of about 60 in neutron flux compared to the today's strongest neutron sources in comparable conditions (e.g. using TOF at the ILL) offers great new opportunities for the field of neutron reflectometry. It will lead to a drastic reduction of the measurement time, allowing for instance a detailed mapping of phase diagrams what is not feasible today. In the field of kinetics, processes in the sub-seconds regime will be accessible which is particularly interesting in many soft matter cases. The increase of neutron flux in combination of focussing techniques will further allow one to investigate smaller samples down to $1mm^2$ sample surface or even smaller.

Another big opportunity inherent with such a drastic increase of the incident neutron flux for neutron reflectometry will be the ability to access lower reflectivities and as a consequence, the extension of the accessible *Q*-range. Data from a larger reciprocal space regime will deliver direct and more accurate information about smaller lengths scales as it is possible today.

Assuming a statistically background level *b* without time dependent fluctuations except of statistical errors and accounting for neutrons scattering obeying the Poisson statistics, we consider only reflectivity signals which have at least 3 times of the fluctuation levels of the background noise to be detectable. This leads to a relationship between the measurement time *t*, the lowest measurable reflectivity level *R* (measured intensity from the specular part of the reflected intensity: $I_S = I_0 R t$, the incident neutron flux $I_0$ (neutrons per sec) and the background level *b* (background intensity: $I_{BG} = I_0 b t$), as follows:

$$R = \sqrt{\frac{b}{I_0} \cdot \frac{9}{t}} \qquad (2)$$

If we consider the ratio of the background level *b* to the incident neutron intensity $I_0$ as a combined parameter and the time as another crucial parameter, the relationship between the lowest necessary counting time in respect to a signal of certain intensity (reflectivity level) can be illustrated as shown in Fig. 22.

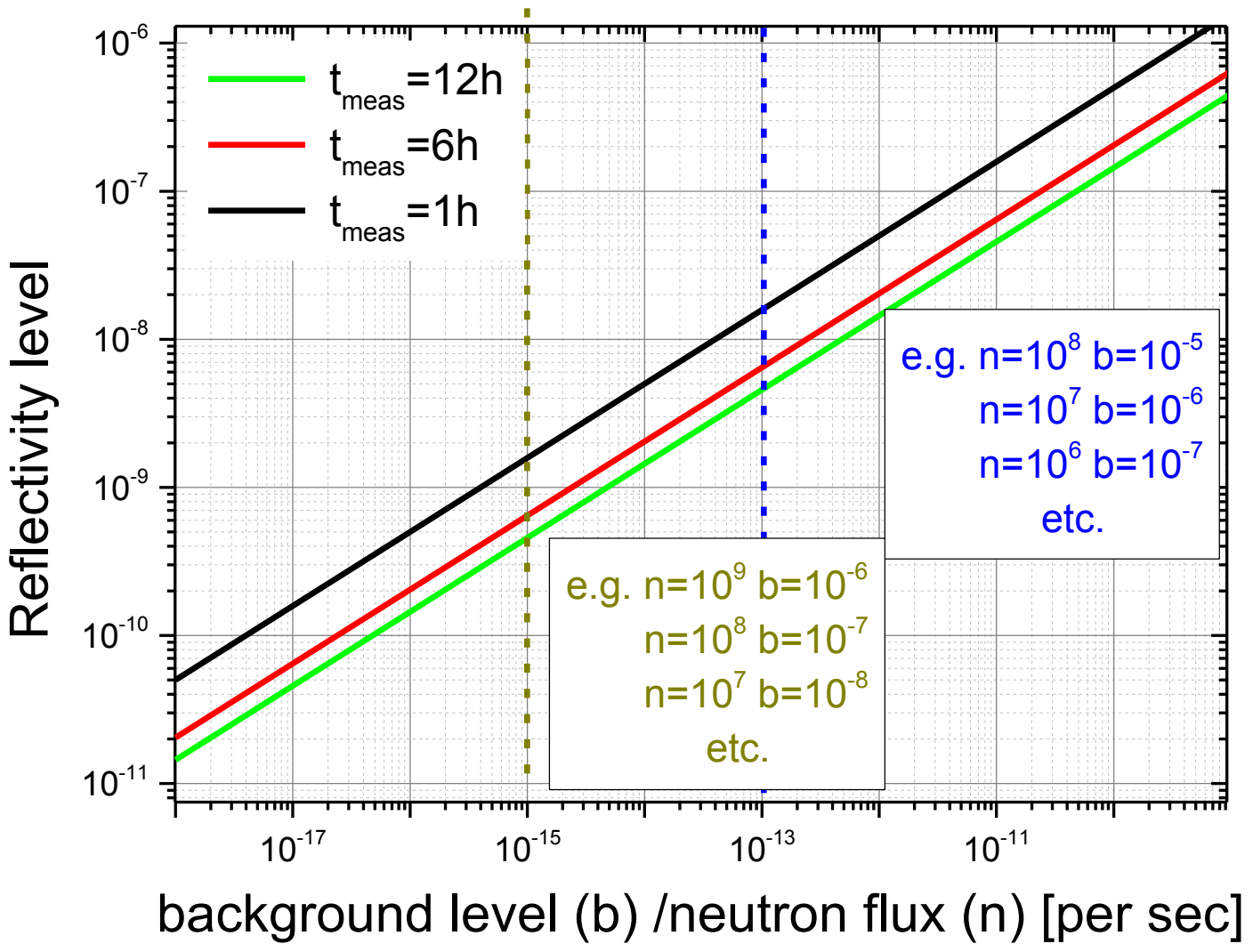

Fig. 22: The coloured full lines indicate the lowest measurable reflectivity level in respect to the relationship between the ratio of the background level *b* to the incident neutron intensity *n* for different measurement time *t*.

By knowing the incident neutron flux and the background level *b*, the lowest reflectivity level can be estimated. With a neutron flux of $10^8$ n/s and a background level of $10^{-7}$, for example, reflectivities of $R=2\cdot10^{-9}$ are already measurable in 1h or reflectivities of $R=5\cdot10^{-10}$ in 12h. The loss in incident neutrons can be direct proportional compensated by a reduction in the background level if one wants to keep the measurement time constant. Having more incident neutrons, on the other hand, allows for a corresponding higher background as long as the ratio between *b* and *n* remains constant.

The considerations here imply that with the huge increase in neutron flux at the ESS and particularly with the HERITAGE concept, it will be possible to measure reflectivities on the order of $10^{-9}$ or even below if the background level can be kept as low as here assumed.

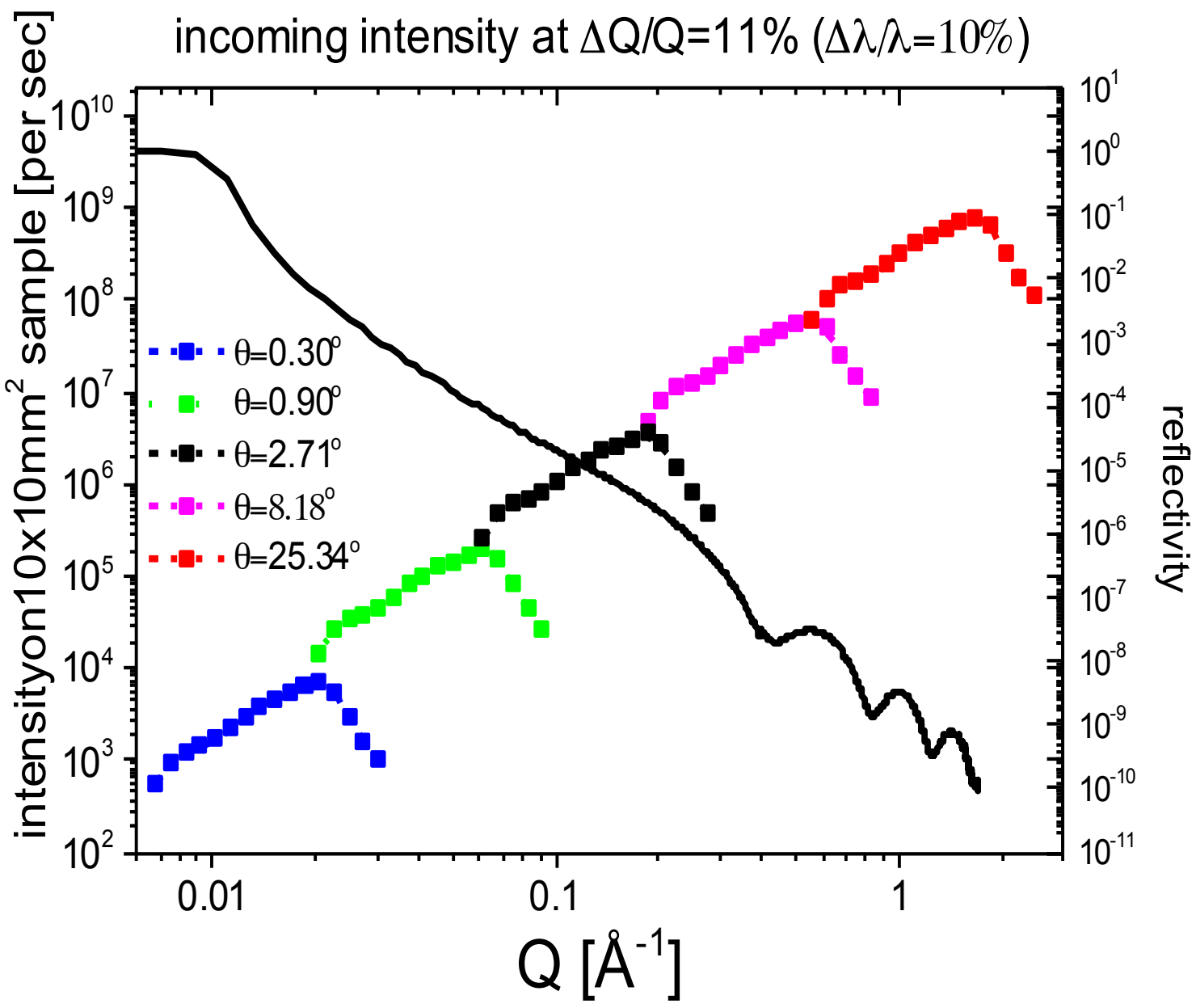

| θ [deg] | Back-ground level | Reflecti-vity level | Time [sec, h] |
|---|---|---|---|
| 0.3 | $6\cdot10^{-7}$ | $1\cdot10^{-2}$ | 1p |
| 0.3 | $1\cdot10^{-5}$ | $1\cdot10^{-2}$ | 1s |
| 0.9 | $6\cdot10^{-7}$ | $1.3\cdot10^{-4}$ | 1p |
| 0.9 | $1\cdot10^{-5}$ | $1.3\cdot10^{-4}$ | 1s |
| 2.71 | $6\cdot10^{-7}$ | $4\cdot10^{-6}$ | 1p |
| 2.71 | $1\cdot10^{-5}$ | $4\cdot10^{-6}$ | 1s |
| 8.18 | $1\cdot10^{-7}$ | $2\cdot10^{-8}$ | 1min |
| 8.18 | $1\cdot10^{-6}$ | $2\cdot10^{-8}$ | 10min |
| 25.34 | $10^{-7}$ | $1\cdot10^{-9}$ | 30min |
| 25.34 | $10^{-6}$ | $1\cdot10^{-9}$ | 5h |
| 25.34 | $10^{-7}$ | $1\cdot10^{-10}$ | 13h |

Fig. 23. Theoretical calculation of the measuring time for a 15Å thick $SiO_2$ layer on Si substrate (without roughness) for different reflectivity levels versus background levels of the real instrument.

In Fig. 23 the theoretical calculations for a sample systems is carried out to illustrate how reflectivity levels of $10^{-9}$ could be measured in five hours or even less. It should be noted, however, that such low reflectivity levels require additional conditions that have to be fulfilled. At first, the incident neutron flux n on the sample is due to sample illumination condition much lower at small as at large incident reflectivity angles. In case of samples with rough interfaces, the reflectivity drops already at very small reflectivity angles to levels where the incident neutron flux is still too low for achieving a reasonable low b/n ratio that enables one to detect the signal. The second point that needs to be paid attention is valid for not only reflectometry, but also all instruments at any neutron sources. It is tremendously crucial that any systematic background errors that are not independently measurable or long term statistical errors that have time scales on the order or longer as the typical measurement time for low reflectivity measurements are avoided since these errors will make it quasi impossible to divide such background from a weak signal. Therefore it is very important that the experimental environment where the instruments will be placed are extremely stable, e.g. for temperature or humidity changes that occur during the day/night cycle or other sources of external fluctuations. Moreover, it will be extremely important that the experiment is very well shielded against parasitical neutrons from measurement of neighbouring instruments since this may be the major source of a non-systematic and non-statistical background contribution.

## 12. Conclusions

The reflectometer concept presented here is not only supporting static and kinetic measurements on all possible types of interfaces (solid to free liquids) but combining unprecedentedly high flux with high lateral resolution allowing for the first time to investigate with one experiment (several measurements) the complete 3d-structure of a thin film sample. The extreme high flux of $7.6 \cdot 10^9$ n/cm$^2$/sec is not only achieved by the outstanding performance of the ESS pancake moderator but it is flanked by the modern neutron guide concept in the horizontal direction focussing on the sample via an elliptical shaped neutron guide. The intensity may be used just for achieving a significant higher dynamic $Q$-range as it is today feasible, for the acceleration of kinetic measurements to time scales that are not accessible today, for a higher $Q$-resolution, for the exploration of 3-d structure in thin films or just for shorter measurement times or measurement of very small samples at a reasonable time.

The strength of the HERITAGE concept lies in its potential to investigate all kind of interfaces types by its ability to quickly switch between the available operational modes inside the collimation by keeping the complexity of the instrument low. Ranging from single and multi-beam setups illuminating the sample from above or below allowing for the measurements of specular, off-specular reflectivity and GISANS combined with the full variety of $\lambda$-resolution, polarisation analysis and kinetic measurements.

The realisation of such a reflectometer concept provides an extremely high flux, vastly exceeding the flux, which is currently achievable at the best reflectometers in the world. Particularly, about 60 times gain is expected in comparison with D17 at the ILL [13] and factor about 90 in comparison with magnetic reflectometer at the SNS [14]. Moreover, in comparison with the planned reflectometer FREIA for ESS, the gain amounts to a factor of about 5 for 4x4cm$^2$ liquid samples and to factor about 8 for gain for 1x1cm$^2$ solid samples.

## Acknowledgements

Authors would like to acknowledge the stimulating discussions with F.Mezei (ESS), R. Cubitt (ILL), Th. Brückel (FZJ), R. Dalgliesh (ISIS), K. Andersen (ESS), D. Argyriou (ESS), P. Müller-Buschbaum (TU Munich), J. Ankner (SNS), R.Pynn (UCSB) and L.Rosta (Wigner Research Center, Budapest).

We also would like to thank S. Manoshin (JINR, Dubna) for the support with VITESS.